\documentclass[aps, prd, showkeys,superscriptaddress, nofootinbib, floatfix]{revtex4-2}

\usepackage{amsmath}
\usepackage{graphicx}
\usepackage{dcolumn}
\usepackage{bm}

\usepackage{amssymb}
\usepackage{dsfont}
\usepackage{url}
\usepackage{caption,subcaption}
\usepackage[colorlinks=true,allcolors=blue]{hyperref}
\newcommand{\appropto}{\mathrel{\vcenter{
  \offinterlineskip\halign{\hfil$##$\cr
    \propto\cr\noalign{\kern2pt}\sim\cr\noalign{\kern-2pt}}}}}

\begin{document}

\title{Relaxation Time Approximation for a multi-species relativistic gas}

\author{Gabriel S. Rocha}
\email{gabrielsr@id.uff.br}
\affiliation{Instituto de F\'{\i}sica, Universidade Federal Fluminense, Niter\'{o}i, Rio de Janeiro, 24210-346,
Brazil}
\affiliation{Department of Physics and Astronomy, Vanderbilt University, Nashville, TN 37240, USA}

\author{Gabriel S. Denicol}
\email{gsdenicol@id.uff.br}
\affiliation{Instituto de F\'{\i}sica, Universidade Federal Fluminense, Niter\'{o}i, Rio de Janeiro, 24210-346,
Brazil}

\begin{abstract}
We generalize a recent prescription for the relaxation time approximation for the relativistic Boltzmann equation for systems with multiple particle species at finite temperature. This is performed by adding counter-terms to the traditional Anderson-Witting ansatz for each particle species. Our approach allows for the use of momentum-dependent relaxation times and the obedience of local conservation laws regardless of the definition of the local equilibrium state. As an application, we derive the first order Chapman-Enskog corrections to the equilibrium distribution and display results for the hadron-resonance gas. We also demonstrate that our collision term ansatz obeys the second law of thermodynamics.
\end{abstract}

\maketitle

\section{Introduction}

The main goal of heavy-ion collisions is to understand the properties of the Quark-Gluon Plasma, a state of nuclear matter in which the fundamental degrees of freedom of Quantum Chromodynamics are deconfined \cite{Harris:2024aov,Ratti:2021ubw}. However, these properties cannot be observed directly since experiments mostly measure particles produced at the late stages of the collision, when the system is already cold and dilute. Thus, it is essential to reconstruct the intermediate steps between the initial deposition of energy by the nuclei and the detection of final state particles, and, for this purpose, several simulation chains have been developed \cite{Putschke:2019yrg,Nijs:2020ors,Hippert:2020kde,JETSCAPE:2020mzn}. 

A crucial step in these multi-stage simulations is the so-called particlization \cite{Huovinen:2012is}, which changes the fluid degrees of freedom, evolved by fluid-dynamical simulations, to particle degrees of freedom 
that will then be evolved using hadronic transport. For this purpose, the interacting hadron gas is approximated as a weakly-interacting hadron-resonance gas (HRG) \cite{Venugopalan:1992hy,Karsch:2003vd,Huovinen:2009yb,Bazavov:2009zn,Borsanyi:2010cj,Vogt:2007zz} and the multiplicity and momenta of the hadrons and resonances are sampled from the single-particle distribution function
using the Cooper-Frye method \cite{Cooper:1974mv,Cooper:1974qi} at a given space-time hypersurface\footnote{This transition is usually assumed to occur at a constant temperature or energy density.}. In local equilibrium, this distribution corresponds to a Fermi-Dirac or Bose-Einstein distribution, depending on the hadronic species. Out-of-equilibrium, determining these momentum distribution exactly is simply not possible yet and additional models and assumptions are often employed \cite{Dusling:2009df,Teaney:2003kp,Pratt:2010jt,Bernhard:2018hnz,Pratt:2010jt}. Recently, it has been noticed that this choice of model can have a significant effect on Bayesian analyses \cite{JETSCAPE:2020mzn,Heffernan:2023utr,Paquet:2023rfd}.

A widely employed method to determine the non-equilibrium correction to the single-particle distribution function is to impose the first order Chapman-Enskog approximation \cite{chapman1916vi,enskog1917kinetische} to the Boltzmann equation in the relaxation time approximation (RTA)\footnote{Other parametrizations for particlization methods include the 14-moments \cite{Dusling:2009df,Teaney:2003kp,Monnai:2009ad,Dusling:2011fd}, the Pratt-Torrieri-Bernhard \cite{Pratt:2010jt,Bernhard:2018hnz}, and the Pratt-Torrieri-McNelis \cite{McNelis:2019auj,Pratt:2010jt} methods.}. 
This approximation imposes that the system approaches local equilibrium exponentially within a characteristic timescale -- the relaxation time.
The traditional Anderson-Witting RTA \cite{andersonRTA:74,ANDERSON1974489} has been extensively used in the field \cite{Bazow:2016oky,Bhadury:2020puc,Kamata:2020mka,Rocha:2022ind,Denicol:2014vaa,Denicol:2014xca,Aniceto:2024pyc,Noronha:2015jia,Tinti:2016bav}, but displays severe limitations: it violates the conservation of energy and momentum if the relaxation time depends on the momentum\footnote{It is expected that realistic approaches to local equilibrium occur with relaxation times that depend on momentum \cite{Denicol:2022bsq,Rocha:2024rce,Calzetta:1986cq,Mukherjee:2025dqp}.} of the particles or if the non-equilibrium prescription for temperature and four-velocity is not that of Landau \cite{landau:59fluid}.  These fundamental flaws stem from the violation of fundamental properties of the collision term in the Anderson-Witting ansatz. In Ref.~\cite{Rocha:2021zcw}, a new RTA ansatz has been put forward where the aforementioned violations are fixed by the addition of counter-terms to the Anderson-Witting RTA \cite{reichl:99,bhatnagar:54model,cercignani:90mathematical}. Throughout the years, other approaches to circumvent the limitations of the Anderson-Witting approach have also been employed \cite{Ambrus:2023ilm,Hu:2022mvl,Dash:2021ibx,Calzetta:2010au,De:2022yxq,Pennisi:18,Carrisi:19,hwang2021relativistic,Bhadury:2025fil}.

In this work, we generalize the results of Ref.~\cite{Rocha:2021zcw} to multi-species systems. The present work is organized as follows: in Sec.~\ref{sec:multi-nRTA} we outline general aspects of kinetic theory for multiple particle species, defining the hydrodynamic variables and the main properties of the collision term. Then, in Sec.~\ref{sec:nRTA-mult-part}, we present the novel RTA prescription for a chargeless system. Afterwards, in Sec.~\ref{sec:1st-order-CE}, we derive the solution of the first order Chapman-Enskog expansion in detail for this collision term Ansatz. There, it is seen that the derivation becomes involved due to the fact that the novel RTA collision term is not trivially invertible: as it happens with the linearized collision term, our RTA collision matrix possess a non trivial null-space which is related to the conservation of four-momentum. The system can only be suitably inverted taking into account the constraints given by the matching conditions, which define the local equilibrium state. As a consistency check, in Section \ref{sec:entropy-production} we demonstrate that the entropy production is indeed non-negative for our RTA Ansatz. After that, in Sec.~\ref{sec:results}, we apply the results of the previous section to the hadron resonance gas and assess the features of a momentum-dependent relaxation time. In the shear sector of the non-equilibrium configurations, a simple extension of the traditional RTA results are seen, whereas the bulk sector contains a novel contribution that is proportional to the energy of the particle. Section \ref{sec:conclusion} concludes the main text, which is complemented by two Appendices. In Appendix \ref{apn:sum-coeffs}, we describe the series summation procedure employed in the first order Chapman-Enskog solution. Then, in Appendix \ref{sec:zero modes and m} we provide details of the fully analytic procedure to invert the novel RTA collision matrix consistently with Landau matching conditions. The main results of this paper are Eqs.~\eqref{eq:del-f-sol-1} and \eqref{eq:fin-del-f-comps}, where we display the first order Chapman-Enskog solution for a generic parametrization of the relaxation time and Eqs.~\eqref{eq:CF1}, \eqref{eq:phi_bulk} and \eqref{eq:phi_shear}, where we show the first order Chapman-Enskog solution for a power-law parametrization of the relaxation time with the energy of the particles. 

{\bf Notation:} In the present work, we employ the mostly minus  $(+,-,-,-)$ metric signature, and natural units so that $\hbar = c = k_{B} = 1$ and in multiple instances the sum over particle species shall be omitted, so that $\sum_{j=1}^{N_{\rm spec}} (\cdots)_{j} = \sum_{j} (\cdots)_{j}$, where $N_{\rm spec}$ is the number of particle species.

\section{Multiple species Boltzmann equations and hydrodynamic variables}
\label{sec:multi-nRTA}

 The description of a dilute multi-component gas is determined by the single-particle momentum distribution function of each particle species $f_{i}(x,p) \equiv f_{{\bf p}, i}$, $i=1, \cdots, N_{\mathrm{spec}}$, where $N_{\mathrm{spec}}$ is the number of particle species\footnote{Alternatively, one may also place the name of the particles in the labels, e.g., $i = \{ \pi^{\pm}, \pi^{0}, p^{\pm}, K^{\pm} , K^{0}, \cdots \}$}. In complete analogy with the single-particle case, the evolution of each function $f_{{\bf p}, i}$ is given by the relativistic Boltzmann equation
\begin{equation}
\label{eq:BEq}
\begin{aligned}
&
p^{\mu}_{i} \partial_{\mu} f_{{\bf p}, i}  =\sum_{j=1}^{N_{\mathrm{spec}}} C_{ij}[f]
\equiv
C_{i}[f],
\end{aligned}    
\end{equation}
where $C_{i}[f]$ is the collision term for the $i$-th particle species, which incorporates all the collision processes between that species and the remaining ones. For instance, for two-to-two processes, the collision matrix $C_{ij}[f]$ is expressed as,
\begin{equation}
\label{eq:collision-term-multi}
\begin{aligned}
&
C_{ij}[f] = \frac{1}{2} \sum_{a,b=1}^{N_{\mathrm{spec}}}  \int dQ_{a} dQ'_{b} dP'_{j} \left( W_{qq' \to pp'}^{ab \to ij} 
f_{{\bf q}, a} f_{{\bf q}', b} \Tilde{f}_{{\bf p}, i}
\Tilde{f}_{{\bf p}', j}
-
W_{pp' \to qq'}^{ij \to ab}
f_{{\bf p}, i} f_{{\bf p}', j} \Tilde{f}_{{\bf q}, a}
\Tilde{f}_{{\bf q}', b}
\right),
\end{aligned}    
\end{equation} 
where $\Tilde{f}_{{\bf p}, i} = 1 - (a_{i}/g_{i}) f_{{\bf p}, i}$, with $a_{i}=0,1,$ and $-1$, respectively, for classical particles, fermions and bosons and $g_i$ being the corresponding degeneracy factor. Moreover, $W_{qq' \to pp'}^{ab \to ij} $ is the transition rate and is given by $W_{qq' \to pp'}^{ab \to ij} = [(2 \pi)^{4}/16] \left\vert \mathcal{M}_{ij \to ab} \right\vert^{2}  \delta^{(4)}(q_{i} + q_{j}' - p_{a} - p_{b}')$, where $\left\vert \mathcal{M}_{ij \to ab} \right\vert^{2}$ is the transition probability \cite{deGroot:80relativistic,Fotakis:2022usk}. We also defined the integral measure for on-shell particles, $dP_{j} = d^{3}p_{j}/[(2 \pi)^{3} p^{0}_{j}]$.

At high collision energies, the net-baryon density of the matter is relatively small. This hinders the observation of any dissipative effects due to baryon diffusion \cite{Monnai:2012jc,Greif:2017byw}. In this regime, a hydrodynamic description of the system considers only the conservation of energy and momentum, 
\begin{equation}
\begin{aligned}
\label{eq:consv-eqns-no-ch}
    \partial_{\mu}T^{\mu \nu} = 0,
\end{aligned}
\end{equation}
where $T^{\mu \nu}$ is the energy-momentum tensor. In kinetic theory, it is determined from the particle distribution functions as, \cite{deGroot:80relativistic}
\begin{equation}
\begin{aligned}
    T^{\mu \nu} = \sum_{i} \int dP_{i} \ p^{\mu}_{i} p^{\nu}_{i} f_{{\bf p}, i},
\end{aligned}
\end{equation}
where we employed the notation $\sum_{i} = \sum_{i=1}^{N_{\rm spec}}$, which shall be used throughout the text. Using the expression above, Eq.~\eqref{eq:consv-eqns-no-ch} can be derived from the Boltzmann equation by integrating both sides of Eq.~\eqref{eq:BEq} with $p^{\mu}_{i}$, summing over particles, and then using the fundamental property of the collision term,
\begin{equation}
\label{eq:property-coll-term}
\begin{aligned}
    \sum_{i} \int dP_{i} \ p^{\mu}_{i} C_{i} [f] =0,
\end{aligned}
\end{equation}
which arises as a consequence of the microscopic conservation of four-momentum.

We also introduce a local equilibrium distribution function, defined in terms of the temperature $T = 1/\beta$ and four-velocity $u^{\mu}$ of the fluid. For the purposes of the present work, we shall consider the local equilibrium state for each species as  
\begin{equation}
\label{eq:local-eql}
\begin{aligned}
&
f_{0{\bf p},i} = \frac{g_{i}}{e^{\beta u_{\mu}p^{\mu}_{i}} + a_{i}}, \quad i = 1, \cdots, N_{\rm spec},
\end{aligned}    
\end{equation}
which can be obtained as the zero-entropy-production single-particle distribution configuration \cite{deGroot:80relativistic}. The meaning of the fields $T$ and $u^{\mu}$ for non-equilibrium configurations is given by the matching conditions \cite{landau:59fluid,Eckart:1940te}. In the present work, we shall consider Landau's prescription \cite{landau:59fluid} where $u^{\mu}$ is defined as an eigenvector of the energy-momentum tensor, $T^{\mu}_{\ \nu} u^{\nu} = \varepsilon_{0} u^{\mu}$. Then, $T$ is defined so that the local energy density, $\varepsilon_{0}$, follows the equilibrium equation of state, i.e., dissipative corrections of the energy density are identically zero. In terms of phase space integrals, Landau matching conditions lead to the constrains
\begin{equation}
\label{eq:matching-kin}
\begin{aligned}
&    
\sum_{i}\int dP_{i} E_{{\bf p},i}^{2} \delta f_{{\bf p},i}
\equiv 
0,
\quad
\sum_{i}\int dP_{i} E_{{\bf p},i} p_{i}^{\langle \mu \rangle} \delta f_{{\bf p},i}
\equiv 
0,
\end{aligned}    
\end{equation}
where $\delta f_{{\bf p},i} \equiv f_{{\bf p},i} - f_{0{\bf p},i}$, $E_{{\bf p},i} = u_{\mu} p^{\mu}_{i}$, and $p_{i}^{\langle \mu \rangle} = \Delta^{\mu}_{\ \nu} p_{i}^{\nu}$ is the space-like projection of the 4-momenta, with $\Delta^{\mu \nu} \equiv g^{\mu \nu} - u^{\mu}u^{\nu}$ being the projector into the subspace orthogonal to $u^{\mu}$.

With the constrains imposed by Landau matching conditions, the most generic decomposition of $T^{\mu \nu}$ in terms of the fluid four-velocity is
\begin{equation}
\begin{aligned}
& T^{\mu \nu}  = \varepsilon_{0} u^{\mu} u^{\nu} - (P_{0}+\Pi) \Delta^{\mu \nu} + \pi^{\mu \nu},    
\end{aligned}    
\end{equation}
where we defined the equilibrium pressure, determined by the equation of state $P_{0} = P_{0}(\varepsilon_{0})$; the bulk viscous pressure, $\Pi$; and the shear-stress tensor, $\pi^{\mu \nu}$. In terms of the phase space integrals of the distribution function, we have
\begin{equation}
\label{eq:def-currents-kin}
\begin{aligned}
&
\varepsilon_{0} = \sum_{i}\int dP_{i} E_{{\bf p},i}^{2} f_{0 {\bf p},i}, 
\quad 
P_{0} = - \frac{1}{3} \sum_{i} \int dP_{i} \left( \Delta_{\mu \nu} p_{i}^{\nu} p_{i}^{\mu} \right) f_{0 {\bf p},i},  \\
&
\Pi = - \frac{1}{3} \sum_{i} \int dP_{i} \ \left( \Delta_{\mu \nu} p_{i}^{\nu} p_{i}^{\mu} \right) \delta f_{{\bf p},i}, 
\quad 
\pi^{\mu \nu} = \sum_{i} \int dP_{i} p^{\langle \mu}_{i} p^{\nu \rangle}_{i} \delta f_{{\bf p},i},
\end{aligned}    
\end{equation}
where $p^{\langle \mu}_{i} p^{\nu \rangle}_{i} = \Delta^{\mu \nu}_{\ \  \alpha \beta} p^{\alpha}_{i} p^{\beta}_{i}$, given in terms of the double-symmetric and traceless tensor $\Delta^{\mu \nu \alpha \beta} = (1/2) \left( \Delta^{\mu \alpha} \Delta^{\nu \beta} + \Delta^{\mu \beta} \Delta^{\nu \alpha} \right) - (1/3) \Delta^{\mu \nu} \Delta^{\alpha \beta}$. 

In the hydrodynamic regime, the collision term can be linearized around the local equilibrium state. For example, for two-to-two processes, this linearization procedure leads to, 
\begin{equation}
\label{eq:collision-term-multi}
\begin{aligned}
C_{i}[f] & \simeq  \frac{1}{2} \sum_{j,a,b=1}^{N_{\mathrm{spec}}}  \int dQ_{a} dQ'_{b} dP'_{j} W_{qq' \to pp'}^{ab \to ij}
f_{0{\bf p}, i} f_{0{\bf p}', j} \Tilde{f}_{0{\bf q}, a}
\Tilde{f}_{0{\bf q}', b}
\left(  
\phi_{{\bf q}, a} + \phi_{{\bf q}', b} - \phi_{{\bf p}, i}
- \phi_{{\bf p}', j} \right) \equiv f_{0{\bf p}, i} \Hat{L}_{i} \phi_{{\bf p}} ,
\\
\phi_{{\bf p},j} & \equiv \frac{\delta f_{{\bf p},j}}{f_{0 {\bf p},j}\Tilde{f}_{0 {\bf p},j}},
\end{aligned}    
\end{equation} 
where we note that the linearized collision term depends on all of the deviation functions, $\phi_{{\bf p},i}$, $i = 1, \cdots, N_{\mathrm{spec}}$. The linearized collision term also possesses special properties with respect to microscopically conserved quantities. It obeys 
\begin{equation}
\label{eq:lin-coll-csv-law}
\begin{aligned}
    \sum_{i} \int dP_{i} \ p^{\mu}_{i} \Hat{L}_{i}\phi_{\bf p} =0,
\end{aligned}
\end{equation}
which is analogous to Eq.~\eqref{eq:property-coll-term} and allows for the derivation of the local conservation laws \eqref{eq:consv-eqns-no-ch} in the linear regime. Furthermore it also obeys the self-adjointness property
\begin{equation}
\label{eq:self-adj}
\begin{aligned}
    \sum_{i} \int dP_{i} \ G_{{\bf p},i} \Hat{L}_{i} H_{{\bf p}}  = \sum_{i} \int dP_{i} \ H_{{\bf p},i} \Hat{L}_{i} G_{{\bf p}}.
\end{aligned}
\end{equation}
where $G_{{\bf p},j}$ and $H_{{\bf p},j}$ are arbitrary functions of momentum and particle species. This implies that the 4-momentum is an eigenvector of $\Hat{L}$ with a vanishing eigenvalue \footnote{The conserved charges are also degenerate eigenvectors of $\Hat{L}$ with vanishing eigenvalue, but they are not relevant to our discussion, since we consider the limit of vanishing local charge.},
\begin{equation}
\label{eq:zero-modes-L}
\begin{aligned}
\Hat{L}_{i} p^{\mu}
=0.
\end{aligned}    
\end{equation}
The linearized collision term is also crucial in the computation of transport properties of hydrodynamic theories, and indeed the transport coefficients in such theories are given in terms of its inverse \cite{deGroot:80relativistic,Denicol:2021,Rocha:2022ind}. Besides, linear response properties of the Boltzmann equation can be assessed thorough the knowledge of $\Hat{L}^{-1}$ \cite{Ochsenfeld:2023wxz, Rocha:2024cge, Gavassino:2024rck, Moore:2018mma}. Nevertheless, the inversion of the linearized collision term poses a challenging task. Indeed, its exact inversion can only be performed in very rare circumstances \cite{Denicol:2022bsq,cercignani:90mathematical,Wagner:2023joq}. The complexity of this inversion procedure has led, throughout the years, to the proposal of phenomenological approximations for $\Hat{L}$, so that analytically accessible results are possible. A notable example of such models is the Relaxation Time Approximation, put forward by Anderson and Witting \cite{andersonRTA:74,ANDERSON1974489}, which is widely used in theoretical developments in Heavy-Ion collision. Next, we shall discuss a version of this approximation for a multi-species system, assess its limitations and propose a new Ansatz inspired in Ref.~\cite{Rocha:2021zcw}.

\section{Novel RTA prescription for a chargeless multi-particle system}
\label{sec:nRTA-mult-part}

Near equilibrium, the (linearized) collision term acts to drive the system towards equilibrium at a rate that is proportional to the deviation of the particle distribution function, $\delta f_{{\bf p},i}$ \cite{reichl:99,cercignani:90mathematical}. Following this argument the Anderson-Witting RTA ansatz for a system of multiple species reads,
\begin{equation}
\label{eq:AW-RTA}
\begin{aligned}
&
f_{0{\bf p}, i} \Hat{L}_{i} \phi_{{\bf p}} \simeq - \frac{E_{{\bf p},i}}{\tau_{R{\bf p},i}} \delta f_{{\bf p},i} = - \frac{E_{{\bf p},i}}{\tau_{R{\bf p},i}} f_{0 {\bf p},i} \Tilde{f}_{0 {\bf p},i} \phi_{{\bf p},i}, 
\end{aligned}    
\end{equation}
where $\tau_{R{\bf p},i}$ is the RTA timescale. We shall consider that $\tau_{R{\bf p},i}$ can depend on the particle species, on the momentum of that given species and on spacetime. In this case, the approximation \eqref{eq:AW-RTA}, however, is not necessarily consistent with the macroscopic conservation laws \eqref{eq:consv-eqns-no-ch}. Indeed, integrating both sides of Eq.~\eqref{eq:BEq} making use of \eqref{eq:AW-RTA} leads to
\begin{equation}
\begin{aligned}
\label{eq:consv-eqns-RTA}
    \partial_{\mu}T^{\mu \nu} = - \sum_{i} \int dP_{i} \frac{E_{{\bf p},i}}{\tau_{R{\bf p},i}} p^{\nu}_{i} \delta f_{{\bf p},i},
\end{aligned}
\end{equation}
which is, in general, not zero. Nevertheless, if Landau matching conditions are used and, the relaxation times, $\tau_{R{\bf p},i}$, do not depend neither on particle momenta nor on any other particle-dependent parameter (e.g.~mass), then $\sum_{i} \int dP_{i} (E_{{\bf p},i}/\tau_{R{\bf p},i}) p^{\mu}_{i} \delta f_{{\bf p},i} = (1/\tau_{R}) (\delta \varepsilon u^{\mu} + h^{\mu}) \equiv 0$, where $\delta \varepsilon \equiv \sum_{i} \int dP_{i} E_{{\bf p},i}^{2} \delta f_{{\bf p},i}$ is the non-equilibrium correction to energy density and $h^{\mu} = \int dP_{i} E_{{\bf p},i} p^{\langle \mu \rangle}_{i} \delta f_{{\bf p},i}$ is energy flux. Both of these quantities are zero by construction for Landau matching conditions (see Eq.~\eqref{eq:matching-kin}). This artificially constrains the scope of the approximation, since for the full Boltzmann equation the conservation laws are valid regardless of the matching conditions and, in principle, there is no reason for the relaxation time to be independent of momentum.  

Schematically, the Anderson-Witting Ansatz \eqref{eq:AW-RTA} can be understood as $\Hat{L} \appropto \mathds{1} $, which obviously violates the fundamental property \eqref{eq:zero-modes-L}. Since the latter stems from the conservation of energy-momentum in microscopic collisions, it is natural to expect that an approximation that violates condition \eqref{eq:zero-modes-L} will lead to a disruption of energy-momentum conservation on a macroscopic level, as seen in Eq.~\eqref{eq:consv-eqns-RTA}. In Ref.~\cite{Rocha:2021zcw}, an approximation to the linearized collision term that does not breach \eqref{eq:property-coll-term} and \eqref{eq:zero-modes-L} was put forward. In order to recover these properties, counter terms are included to guarantee the validity of the conservation laws.  The approximation can be understood as \cite{Rocha:2021zcw,Rocha:2022fqz}
\begin{equation}
\begin{aligned}
\Hat{L} \hspace{-0.3cm} \appropto \hspace{-0.3cm} - \frac{\mathds{1}}{\tau_{R}} +  \frac{1}{\tau_{R}} \sum_{n} \vert \mathcal{Q}_{n,{\bf p}} \rangle \langle \mathcal{Q}_{n,{\bf p}} \vert,     
\end{aligned}    
\end{equation}
where $\{\mathcal{Q}_{n,p}\}$ is an orthogonal basis constructed from the microscopically conserved quantities. In the present case, only the components of the 4-momentum, $p^{\mu}_{i}$, are considered. In practice, this leads to the following Ansatz 
\begin{equation}
\label{eq:NRTA}
\begin{aligned}
f_{0{\bf p}, i} \Hat{L}_{i} \phi_{{\bf p}}  \simeq - \frac{E_{{\bf p},i}}{\tau_{R{\bf p},i}} f_{0 {\bf p},i} \Tilde{f}_{0 {\bf p},i} \left\{ \phi_{{\bf p},i}  
-
\frac{\sum_{j} \langle \phi_{{\bf p},j} , E_{{\bf p},j} \rangle}{\sum_{j} \langle E_{{\bf p},j} , E_{{\bf p},j} \rangle} E_{{\bf p},i} 
-
\frac{\sum_{j} \langle \phi_{{\bf p},j} , p_{j}^{\langle \mu \rangle} \rangle}{ \frac{1}{3} \sum_{j} \langle p_{j}^{\langle \nu \rangle} ,  p_{j\langle \nu \rangle} \rangle} p_{i\langle \mu \rangle}   \right\},
\end{aligned}
\end{equation}
where we define the relaxation-time-dependent inner product 
\begin{equation}
\begin{aligned}
 &
\langle G_{{\bf p},j} , H_{{\bf p},j} \rangle = \int dP_{j} \frac{E_{{\bf p},j}}{\tau_{R{\bf p},j}} G_{{\bf p},j}  H_{{\bf p},j} f_{0 {\bf p},j} \Tilde{f}_{0 {\bf p},j}. 
\end{aligned}    
\end{equation}
where $G_{{\bf p},j}$ and $H_{{\bf p},j}$ are once again arbitrary function of the momentum of particle $j$. \textcolor{black}{We remark that the novel relaxation time approximation constructed above for multi-species gases has been also extended to include the effects of \textit{quantum statistics}, while in Ref.~\cite{Rocha:2021zcw} it was originally developed for single-component \textit{classical} particles \footnote{The inclusion of quantum statistics in the novel relaxation time approximation for a single-species gas was developed in Ref.~\cite{Mukherjee:2025dqp}}.}

It is also noted that in Eq.~\eqref{eq:NRTA} the approximation for the collision term contains contribution from all of the other particles, something not seen in Eq.~\eqref{eq:AW-RTA}. To explicitly show that the conservation laws \eqref{eq:consv-eqns-no-ch} are indeed valid for the novel RTA, we integrate Eq.~\eqref{eq:BEq} with $p^{\nu}_{i}$ and sum in $i$ making use of Eq.~\eqref{eq:NRTA}, which yields  
\begin{equation}
\begin{aligned}
\partial_{\mu}T^{\mu \nu} &= - \sum_{i} \left\langle p^{\nu}_{i} , \phi_{{\bf p},i} \right\rangle 
+
\frac{\sum_{j} \langle \phi_{{\bf p},j} , E_{{\bf p},j} \rangle}{\sum_{j} \langle E_{{\bf p},j} , E_{{\bf p},j} \rangle} \sum_{i} \left\langle p^{\nu}_{i} , E_{{\bf p},i} \right\rangle  
+
\frac{\sum_{j} \langle \phi_{{\bf p},j} , p_{j}^{\langle \mu \rangle} \rangle}{ \frac{1}{3} \sum_{j} \langle p_{j}^{\langle \alpha \rangle} ,  p_{j\langle \alpha \rangle} \rangle} \sum_{i} \left\langle p^{\nu}_{i} , p_{i\langle \mu \rangle} \right\rangle  \\
&
= \left(- \sum_{i} \left\langle E_{{\bf p},i} , \phi_{{\bf p},i} \right\rangle 
+
\frac{\sum_{j} \langle \phi_{{\bf p},j} , E_{{\bf p},j} \rangle}{\sum_{j} \langle E_{{\bf p},j} , E_{{\bf p},j} \rangle} \sum_{i} \left\langle E_{{\bf p},i} , E_{{\bf p},i} \right\rangle  
+
\frac{\sum_{j} \langle \phi_{{\bf p},j} , p_{j}^{\langle \mu \rangle} \rangle}{ \frac{1}{3} \sum_{j} \langle p_{j}^{\langle \nu \rangle} ,  p_{j\langle \nu \rangle} \rangle} \sum_{i} \left\langle E_{{\bf p},i} , p_{i\langle \mu \rangle} \right\rangle \right) u^{\nu} \\
&
-
\sum_{i} \left\langle p^{\langle \nu \rangle}_{i} , \phi_{{\bf p},i} \right\rangle 
+
\frac{\sum_{j} \langle \phi_{{\bf p},j} , E_{{\bf p},j} \rangle}{\sum_{j} \langle E_{{\bf p},j} , E_{{\bf p},j} \rangle} \sum_{i} \left\langle p^{\langle \nu \rangle}_{i} , E_{{\bf p},i} \right\rangle  
+
\frac{\sum_{j} \langle \phi_{{\bf p},j} , p_{j}^{\langle \mu \rangle} \rangle}{ \frac{1}{3} \sum_{j} \langle p_{j}^{\langle \alpha \rangle} ,  p_{j\langle \alpha \rangle} \rangle} \sum_{i} \left\langle p^{\langle \nu \rangle}_{i} , p_{i\langle \mu \rangle} \right\rangle
= 0
\end{aligned}    
\end{equation}
where in the second equality the identity $p^{\mu}_{i} = E_{{\bf p},i} u^{\mu} + p^{\langle \mu \rangle}_{i}$ has been employed, and in the third,
$\left\langle E_{{\bf p},i} , p^{(i)}_{\langle \mu \rangle} \right\rangle = 0$ and $\left\langle p^{\langle \mu \rangle}_{i} , p^{\langle \nu \rangle}_{i} \right\rangle = (1/3) \left\langle p^{\langle \alpha \rangle}_{i} , p^{(i)}_{\langle \alpha \rangle} \right\rangle \Delta^{\mu \nu}$. The above result is independent of the specific parametrization of the relaxation time with respect to energy and particle species. In the following sections, we employ Eq.~\eqref{eq:NRTA} to derive non-equilibrium corrections to the equilibrium distribution function, which are used in particlization models.

\section{First order Chapman-Enskog expansion}
\label{sec:1st-order-CE}

The hydrodynamic regime is characterized by a separation between the microscopic scales associated with particle interactions and the characteristic macroscopic scales of the system. In this regime, the non-equilibrium dynamics can be treated perturbatively, using the traditional Chapman-Enskog expansion \cite{deGroot:80relativistic,cercignani:90mathematical,Denicol:2021}. In this procedure, one inserts a book-keeping perturbative parameter $\epsilon$ on the left-hand side of the Boltzmann equation, which, using the Ansatz \eqref{eq:NRTA}, reads
\begin{equation}
\label{eq:chap-ensk0-no-ch}
\begin{aligned}
\epsilon\left( E_{{\bf p},i} D f_{{\bf p},i} + p^{\mu}\nabla_{\mu} f_{{\bf p},i} \right) = - \frac{E_{{\bf p},i}}{\tau_{R{\bf p},i}} f_{0 {\bf p},i} \Tilde{f}_{0 {\bf p},i} \left\{ \phi_{{\bf p},i}  
-
\frac{\sum_{j} \langle \phi_{{\bf p},j} , E_{{\bf p},j} \rangle}{\sum_{j} \langle E_{{\bf p},j} , E_{{\bf p},j} \rangle} E_{{\bf p},i} 
-
\frac{\sum_{j} \langle \phi_{{\bf p},j} , p_{j}^{\langle \mu \rangle} \rangle}{ \frac{1}{3} \sum_{j} \langle p_{j}^{\langle \nu \rangle} ,  p_{j\langle \nu \rangle} \rangle} p_{i\langle \mu \rangle}   \right\},
\end{aligned}
\end{equation}
where we have used that $\partial_{\mu} = u_{\mu} D + \nabla_{\mu}$. One then considers an asymptotic series solution for the single particle distribution function and for its time-like derivative,
\begin{equation}
\label{eq:chap-ensk-1-o1}
\begin{aligned}
f_{{\bf p},i} & =  \sum_{k=0}^{\infty} \epsilon^{k} f^{(k)}_{{\bf p},i}, \\
D f_{{\bf p},i} &= \sum_{k=0}^{\infty} \epsilon^{k} [D f_{{\bf p},i}]^{(k)}.
\end{aligned}
\end{equation}
The parameter $\epsilon$ represents the ratio between the characteristic scale of the collision processes (e.g.~the mean-free path) and the characteristic spacetime scale over which the thermodynamic fields change. The expansion of the time-like derivative in $\epsilon$ is rooted in the conservation laws, which relate all time-like derivatives of the hydrodynamic fields to space-like ones, at a given order \cite{cercignani:02relativistic,Rocha:2022ind}. 

Solving Eq.~\eqref{eq:chap-ensk0-no-ch} at zero-th power in $\epsilon$, we have that the collision term should vanish. Since a vanishing collision term leads to a vanishing entropy production and local equilibrium is the only configuration with zero entropy production, Eq.~\eqref{eq:local-eql} is the zero-th order solution, i.e.~$f_{{\bf p},i}^{(0)} =f_{0{\bf p},i}$. At first order, one must collect all the zero-th order terms in the left-hand side of Eq.~\eqref{eq:chap-ensk0-no-ch}, 
\begin{equation}
\label{eq:chap-ensk-1st}
\begin{aligned}
&  E_{{\bf p},i} [D f_{{\bf p},i}]^{(0)} + p^{\mu}\nabla_{\mu} f_{{\bf p},i}^{(0)}  = - \frac{E_{{\bf p},i}}{\tau_{R{\bf p},i}} f_{0 {\bf p},i} \Tilde{f}_{0 {\bf p},i} \left\{ \phi_{{\bf p},i}^{(1)}  
-
\frac{\sum_{j} \langle \phi_{{\bf p},j}^{(1)} , E_{{\bf p},j} \rangle}{\sum_{j} \langle E_{{\bf p},j} , E_{{\bf p},j} \rangle} E_{{\bf p},i} 
-
\frac{\sum_{j} \langle \phi_{{\bf p},j}^{(1)} , p_{j}^{\langle \mu \rangle} \rangle}{ \frac{1}{3} \sum_{j} \langle p_{j}^{\langle \nu \rangle} ,  p_{j\langle \nu \rangle} \rangle} p_{i\langle \mu \rangle}   \right\},    
\end{aligned}    
\end{equation}
where we defined,
\begin{equation}
\begin{aligned}
\phi_{{\bf p},i}^{(1)} & = \frac{f^{(1)}_{{\bf p},i}}{f_{0 {\bf p},i} \Tilde{f}_{0 {\bf p},i}}.
\end{aligned}    
\end{equation}
We remark that in the Chapman-Enskog procedure one cannot identify $[D f_{{\bf p},i}]^{(0)}$ as $ D f_{{\bf p},i}^{(0)}$. These quantities are nevertheless related. In fact,
\begin{equation}
\begin{aligned}
D f_{{\bf p},i}^{(0)} 
& =
\left[
- E_{{\bf p},i}^{2} D\beta
- 
\beta Du_{\mu} p^{\langle \mu \rangle}_i
\right]f_{0{\bf p},i} \Tilde{f}_{0{\bf p},i}
=
\left[ - \beta E_{{\bf p},i} c_{s}^{2}  \theta   -  \frac{\beta}{\varepsilon_{0} + P_{0}} p^{\langle \mu \rangle}_{i} \nabla_{\mu}P_{0}\right] f_{0{\bf p},i} \Tilde{f}_{0{\bf p},i}
+
\mathcal{O}(\Pi, \pi^{\mu \nu}),
\end{aligned}
\end{equation}
where in the last equality conservation laws have been employed to replace the time-like derivatives with the corresponding space-like ones. All dissipative contributions, which are at least of $\mathcal{O}(\epsilon)$, are collected in the term $\mathcal{O}(\Pi, \pi^{\mu \nu})$. Then, we identify
\begin{equation}
\label{eq:time-like-D-subst}
\begin{aligned}
[D f_{{\bf p},i}]^{(0)}
& =
\left[ - \beta E_{{\bf p},i} c_{s}^{2}  \theta   -  \frac{\beta}{\varepsilon_{0} + P_{0}} p^{\langle \mu \rangle}_{i} \nabla_{\mu}P_{0}\right] f_{0{\bf p},i} \Tilde{f}_{0{\bf p},i},
\\
c_{s}^{2} &= \frac{\partial P_{0}}{\partial \varepsilon_{0}} = \frac{\mathcal{J}_{3,1}}{\mathcal{J}_{3,0}}, 
\end{aligned}    
\end{equation}
where we define the following thermodynamic integrals
\begin{subequations}
\label{eq:thermodynamic_integrals}
\begin{align}
\label{eq:thermodynamic_integrals-I}
\mathcal{I}_{n,q} &= \sum_{i}  I_{n,q}^{(i)}, \\ 
\label{eq:thermodynamic_integrals-J}
\mathcal{J}_{n,q} &= \sum_{i}  J_{n,q}^{(i)},
\\ 
\label{eq:thermodynamic_integrals-Ii}
 I_{n,q}^{(i)} &= \frac{1}{(2q+1)!!} \left\langle \left( -\Delta^{\lambda \sigma}p_{\lambda} p_{\sigma}^{(i)} \right)^{q} E_{\mathbf{p},i}^{n-2q} \right\rangle_{\mathrm{eq},i},
 \\
 \label{eq:thermodynamic_integrals-Ji}
 J_{n,q}^{(i)} &= \frac{1}{(2q+1)!!} \left\langle \left( -\Delta^{\lambda \sigma}p_{\lambda} p_{\sigma}^{(i)} \right)^{q} E_{\mathbf{p},i}^{n-2q} \right\rangle_{\widetilde{\mathrm{eq}},i}. 
\end{align}
\end{subequations}
Above, we employed the following notation
\begin{equation}
\begin{aligned}
\langle (\cdots) \rangle_{\mathrm{eq},i} & = \int dP_{i} (\cdots) f_{0{\bf p},i},
\\
\langle (\cdots) \rangle_{\widetilde{\mathrm{eq}},i}
&
=
\int dP_{i} (\cdots) f_{0{\bf p},i}\Tilde{f}_{0{\bf p},i}.
\end{aligned}    
\end{equation}
Using that $\varepsilon_{0} = \mathcal{I}_{2,0}$ and $P_{0} = \mathcal{I}_{3,1}$, we can define the entropy density using the following thermodynamic relation $\varepsilon_{0} + P_{0} = T s_{0} = T \partial P_{0}/\partial T = \mathcal{J}_{3,1}/T$. 

The first order correction to the equilibrium distribution is determined from the following inhomogenous linear integral equation,
\begin{subequations}
\label{eq:chap-ensk-1st-ord}
\begin{align}
\label{eq:chap-ensk-1st-ord-a}
\left[ \mathcal{A}_{{\bf p},i} \theta  - \beta p^{\langle \mu}_{i}p^{\nu \rangle}_{i} \sigma_{\mu \nu} \right]f_{0 {\bf p},i} \Tilde{f}_{0 {\bf p},i}  & = - \frac{E_{{\bf p},i}}{\tau_{R{\bf p},i}} f_{0 {\bf p},i} \Tilde{f}_{0 {\bf p},i} \left\{ \phi_{{\bf p},i}^{(1)}  
-
\frac{\sum_{j} \langle \phi_{{\bf p},j}^{(1)} , E_{{\bf p},j} \rangle}{\sum_{j} \langle E_{{\bf p},j} , E_{{\bf p},j} \rangle} E_{{\bf p},i} 
-
\frac{\sum_{j} \langle \phi_{{\bf p},j}^{(1)} , p_{j}^{\langle \mu \rangle} \rangle}{ \frac{1}{3} \sum_{j} \langle p_{j}^{\langle \nu \rangle} ,  p_{j\langle \nu \rangle} \rangle} p_{i\langle \mu \rangle}   \right\},
\\
\label{eq:chap-ensk-1st-ord-b}
\mathcal{A}_{{\bf p},i}  &= - \beta E_{{\bf p},i}^{2} c_{s}^{2} - \frac{\beta}{3} \left( \Delta_{\mu \nu} p_{i}^{\nu} p_{i}^{\mu} \right).
\end{align}
\end{subequations}
We note that $\mathcal{A}_{{\bf p},i}$ is orthogonal to the 4-momentum,
\begin{equation}
\label{eq:AE-zero}
\begin{aligned}
\sum_{i} \left\langle \mathcal{A}_{{\bf p},i} p^\mu_i\right\rangle_{\widetilde{\mathrm{eq}},i} = 0. \\
\end{aligned}   
\end{equation}

We now expand $\phi_{{\bf p},i}^{(1)}$ using a complete and orthogonal basis in momentum space, 
\begin{equation}
\label{eq:expn-tensorial}
\begin{aligned}
\phi^{(1)}_{{\bf p},i} = \sum_{\ell,n = 0}^{\infty}  \Phi_{i,n}^{\mu_{1} \cdots \mu_{\ell}} P_{{\bf p},i,n}^{(\ell)} p_{i\langle \mu_{1}} \cdots  p_{i\mu_{\ell} \rangle}, \quad \quad i = 1, \cdots, N_{\mathrm{spec}},
\end{aligned}
\end{equation}
where the basis is defined in terms of the irreducible tensors $p_{i}^{\langle \mu_{1}} \cdots p^{\mu_{\ell} \rangle}_{i} \equiv \Delta^{\mu_{1} \cdots \mu_{\ell}}_{\ \ \nu_{1} \cdots \nu_{\ell}} p^{\nu_{1}}_{i} \cdots p^{\nu_{\ell}}_{i}$ which are constructed from the 4-momentum of each particle species, with $\Delta^{\mu_{1} \cdots \mu_{\ell}}_{\nu_{1} \cdots \nu_{\ell}}$ being the $2\ell$--rank projection operator orthogonal to the fluid 4-velocity in every index. Such projection tensors are constructed from the projectors $\Delta^{\mu\nu}=g^{\mu\nu} - u^{\mu}u^{\nu}$ in such a way that they are traceless in each subset of indices ($\mu_{1} \cdots \mu_{\ell}$) and ($\nu_{1} \cdots \nu_{\ell}$), for $\ell > 1$ and symmetric under the exchange of the indices in each subset, separately \cite{deGroot:80relativistic,Denicol:2021}. The main property of such tensors is the orthogonality condition,
\begin{equation}
\label{eq:orth-tens-ell}
\begin{aligned}
\int dP
p^{\langle \mu_{1}}_{i} \cdots p^{\mu_{\ell} \rangle}_{i}
p_{i\langle \nu_{1}} \cdots p_{i\nu_{m} \rangle}
H(E_{\mathbf{p}})
& = \frac{\ell!\delta_{\ell m}}{(2\ell + 1)!!}  \Delta^{\mu_{1} \cdots \mu_{\ell}}_{\nu_{1} \cdots \nu_{\ell}} \int dP \left(\Delta_{\mu \nu} p_{i}^{\mu} p_{i}^{\nu} \right)^{\ell} H(E_{\mathbf{p}}),   
\end{aligned}    
\end{equation}
where $H(E_{\mathbf{p}})$ is an arbitrary weight function so that the integral on the right-hand side converges. Besides, in the definition of the expansion basis in Eq.~\eqref{eq:expn-tensorial}, we define a set of orthogonal polynomials $P_{{\bf p},i,n}^{(\ell)} \equiv P_{i,n}^{(\ell)}(\beta E_{\bf p})$. For a given species $i$, their orthogonality is given with respect to the inner product
\begin{equation}
\label{eq:orth-polyn-ell}
\begin{aligned}
\left\langle \left( \Delta_{\mu \nu} p_{i}^{\mu} p_{i}^{\nu} \right)^{\ell} \frac{E_{{\bf p},i}}{\tau_{R {\bf p},i}}  P^{(\ell)}_{{\bf p},i,n} P^{(\ell)}_{{\bf p}, i, m} \right\rangle_{\widetilde{\mathrm{eq}},i} = A^{(\ell)}_{n,i} \delta_{mn}. 
\end{aligned}
\end{equation}
In the particular case of massless, classical ($a_{i} = 0$ in Eq.~\eqref{eq:local-eql}) particles, we have that the orthogonal polynomials reduce to associated Laguerre polynomials \cite{NIST:DLMF}. In more general cases, they must be constructed by, e.g., Gram-Schmidt orthogonalization algorithm. We remark that our final expression for the single-particle distribution function will not display an explicit dependence on these polynomials. It is only important to notice that, due to the form of novel RTA the collision term \eqref{eq:NRTA}, it is convenient to start the Gram-Schmidt process with $P^{(0)}_{{\bf p},i,0} = \beta E_{{\bf p},i}$ (for $\ell = 0$) and with $P^{(\ell)}_{{\bf p},i,0} = 1$, for $\ell = 1,2, \cdots$ \cite{Rocha:2022fqz}. 

Substituting expansion \eqref{eq:expn-tensorial} in Eq.~\eqref{eq:chap-ensk-1st-ord}, we have
\begin{equation}
\label{eq:CRTAs}
\begin{aligned}
\left[ \mathcal{A}_{{\bf p},i} \theta - \beta p^{\langle \mu}_{i}p^{\nu \rangle}_{i} \sigma_{\mu \nu} \right]f_{0{\bf p},i}\Tilde{f}_{0 {\bf p},i} = 
\sum_{\ell = 0}^{\infty} C_{{\rm RTA},\ell}  ,
\end{aligned}    
\end{equation}
where we use the notation,
\begin{subequations}
\label{sec:coll-l=0}
\begin{align}
C_{{\rm RTA},0} &= - \frac{E_{{\bf p},i}}{\tau_{R{\bf p},i}} f_{0 {\bf p},i} \Tilde{f}_{0 {\bf p},i} \left\{ 
\sum_{j} \left( \delta_{ij}  - \frac{A^{(0)}_{j,0}}{ \sum_{k} A^{(0)}_{k,0} }\right) \Phi_{j,0} \beta E_{{\bf p},i} 
+
\sum_{n = 1}^{\infty}  \Phi_{i,n} P_{{\bf p},i,n}^{(0)}  \right\} ,
\\
\label{sec:coll-l=1}
C_{{\rm RTA},1} &= -\frac{E_{{\bf p},i}}{\tau_{R{\bf p},i}} f_{0{\bf p},i} \Tilde{f}_{0 {\bf p},i} \left\{ 
\sum_{j} \left( \delta_{ij}  - \frac{A^{(1)}_{j,0}}{ \sum_{k} A^{(1)}_{0,k} }\right) \Phi^{\mu}_{j,0}  p_{i\langle \mu \rangle}  
+
\sum_{n = 1}^{\infty}  \Phi_{i,n}^{\mu} P_{{\bf p},i,n}^{(1)} p_{i\langle \mu \rangle} 
\right\}
,
\\
\label{sec:coll-l>=2}
C_{{\rm RTA},\ell \geq 2} & =
- \frac{E_{{\bf p},i}}{\tau_{R{\bf p},i}} f_{0{\bf p},i} \Tilde{f}_{0 {\bf p},i} \sum_{\ell \geq 2,n = 0}^{\infty}  \Phi_{i,n}^{ \mu_{1} \cdots \mu_{\ell} } P_{{\bf p},i,n}^{(\ell)} p_{i\langle \mu_{1}} \cdots  p_{i\mu_{\ell} \rangle}.
\end{align}    
\end{subequations}
Integrating both sides of the above expression with $P_{{\bf p},i,n}^{(\ell)} p_{i\langle \mu_{1}} \cdots  p_{i\mu_{\ell} \rangle}$ and using the orthogonality properties \eqref{eq:orth-tens-ell} and \eqref{eq:orth-polyn-ell}, we can readily solve Eq.~\eqref{eq:CRTAs} for most of the $\Phi_{i,n}^{\mu_{1} \cdots \mu_{\ell}}$ components,
\begin{subequations}
\label{eq:soln-Phis}
\begin{align}
\label{eq:soln-Phis-l=0}
\Phi_{i,n} &=  - \frac{\theta}{A_{i,n}^{(0)}} \left\langle \mathcal{A}_{{\bf p},i} P^{(0)}_{{\bf p},i,n} \right\rangle_{\widetilde{\mathrm{eq}},i}, \quad \forall i = 1, \cdots, N_{\mathrm{spec}}, \quad  \forall n = 1, 2, \cdots, \\
\label{eq:soln-Phis-l=1}
\Phi_{i,n}^{\mu} &=  0, \quad \forall i = 1, \cdots, N_{\mathrm{spec}}, \quad  \forall n = 1, 2, \cdots,\\
\label{eq:soln-Phis-l=2}
\Phi_{i,n}^{\mu \nu} &=  \beta \frac{\sigma^{\mu \nu}}{A_{i,n}^{(2)}} \left\langle \left(\Delta_{\alpha \beta} p_{i}^{\alpha} p_{i}^{\beta}\right)^{2}  P^{(2)}_{{\bf p},i,n} \right\rangle_{\widetilde{\mathrm{eq}},i}, \quad \forall i = 1, \cdots, N_{\mathrm{spec}}, \quad \forall n = 0, 1, \cdots, \\
\label{eq:soln-Phis-l>2}
\Phi_{i,n}^{\mu_{1} \cdots \mu_{\ell}} &= 0, \quad \forall i = 1, \cdots, N_{\mathrm{spec}}, \quad \forall n = 0, 1, \cdots, \quad \forall \ell = 3, 4, \cdots.
\end{align}
\end{subequations}
The $\Phi_{i,n}^{\mu_{1} \cdots \mu_{\ell}}$ components which cannot be trivially obtained through this manner, $\Phi_{i,0}$ and $\Phi_{i,0}^{\mu}$, for $i = 1, \cdots, N_{\mathrm{spec}}$, are the ones associated with $P^{(0)}_{{\bf p},0,i} = \beta E_{{\bf p},i}$ and $P^{(1)}_{{\bf p},0,i}p^{\langle \mu \rangle}_{i} = p^{\langle \mu \rangle}_{i}$, respectively (cf.~Eq.~\eqref{eq:expn-tensorial}). This is expected because they are the zero-modes of the linearized collision term (see Eq.~\eqref{eq:zero-modes-L}). The components $\Phi_{i,0}$ and $\Phi_{i,0}^{\mu}$ shall be fixed by the matching conditions. However, since the matching conditions involve sums over all particle species and the novel RTA collision term mixes information from different species in a non-trivial way, the results will be more convoluted than in the single particle case.

At this point it is useful to collect the results obtained so far. The partial solution for the deviation function can be expressed as 
\begin{equation}
\label{eq:phi-part-sol}
\begin{aligned}
&
\phi_{{\bf p},i}^{(1)} 
=
\left. \Hat{\phi}^{(1)}_{{\bf p},i} \right\vert_{\ell = 0}
+
\Phi_{i,0}^{\mu} p_{i\langle \mu \rangle}
+
F^{(2)}_{\mathbf{p},i} \sigma_{\mu \nu} \ p^{\langle \mu}_{i} p^{\nu \rangle}_{i} ,
\quad \forall i = 1, \cdots, N_{\mathrm{spec}},
\end{aligned}    
\end{equation}
where the functions $\left. \Hat{\phi}^{(1)}_{{\bf p},i} \right\vert_{\ell = 0}$ and $\left. \Hat{\phi}^{(1)}_{{\bf p},i} \right\vert_{\ell = 2}$ are given by,
\begin{subequations}
\label{eq:phi-all-sol}
\begin{align}
\label{eq:phi-all-sol-scalar}
\left. \Hat{\phi}^{(1)}_{{\bf p},i} \right\vert_{\ell = 0}
& =
\Phi_{i,0} P_{{\bf p},i,0}^{(0)}
+
\left(
- \frac{\tau_{R{\bf p},i}}{E_{{\bf p},i}}  \mathcal{A}_{{\bf p},i}
+
\frac{1}{A^{(0)}_{i,0}} \left\langle  \mathcal{A}_{{\bf p},i} P^{(0)}_{{\bf p},i,0} \right\rangle_{\widetilde{\mathrm{eq}},i} P^{(0)}_{{\bf p},i,0}
\right) \theta, 
\\
\label{eq:phi-all-sol-tensor}
F^{(2)}_{\mathbf{p},i}
& = \beta \frac{\tau_{R{\bf p},i}}{E_{{\bf p},i}},
\end{align}    
\end{subequations}
where one is reminded that $P^{(0)}_{{\bf p},i,0} = \beta E_{{\bf p}, i}$, and the details of the derivation are in Appendix \ref{apn:sum-coeffs}, following the procedure first outlined in Ref.~\cite{Rocha:2022fqz,Rocha:2021zcw}.

Now we turn our attention to obtaining the expansion coefficients $\Phi_{i,0}$ and $\Phi_{i,0}^{\mu}$ $(i = 1, \cdots, N_{\mathrm{spec}})$ by inverting the novel RTA collision matrix consistently with Landau matching conditions. Then, we multiply Eq.~\eqref{eq:CRTAs}  with $P_{i,0}^{(0)} = \beta E_{{\bf p},i}$ and with $p^{\langle \mu \rangle}_{i}$, and integrate them in momentum. This leads to the following constrains, 
\begin{subequations}
\label{eq:sys-0-mode}
\begin{align}
&
\label{eq:sys-0-mode-l=0}
\sum_{j} \mathcal{M}^{(0)}_{ij} \Phi_{j,0} = - \langle \mathcal{A}_{{\bf p},i} P_{{\bf p},i,0}^{(0)} \rangle \theta, \quad \sum_{j} \mathcal{M}^{(1)}_{ij}\Phi^{\mu}_{j,0} = 0,\\
\label{eq:ML-def}
&
\mathcal{M}^{(\ell)}_{ij}
=
A^{(\ell)}_{i,0}\left( \delta_{ij}  - \frac{A^{(\ell)}_{j,0}}{ \sum_{k} A^{(\ell)}_{k,0} }\right), \quad \ell = 0,1.
\end{align}    
\end{subequations}
Naively, since the above expressions form linear systems of $N_{\rm spec}$ equations for the $N_{\rm spec}$ unknowns $\Phi_{i,0}, \ i = 1, \cdots, N_{\rm spec}$ (and $\Phi_{i,0}^{\mu}, \ i = 1, \cdots, N_{\rm spec}$, respectively), we would solve them by simply inverting the symmetric matrices $\mathcal{M}^{(0)}$ and $\mathcal{M}^{(1)}$. However, property \eqref{eq:lin-coll-csv-law}, that is also observed for the novel RTA collision term, implies that the matrices $\mathcal{M}^{(0)}_{ij}$, $\mathcal{M}^{(1)}_{ij}$ obey
\begin{equation}
\label{eq:prop-M0-M1}
\begin{aligned}
&
\sum_{i} \mathcal{M}^{(0)}_{ij} = 0 = \sum_{i} \mathcal{M}^{(1)}_{ij}, 
\end{aligned}    
\end{equation}
This implies that $\det \mathcal{M}^{(0)} = 0 = \det \mathcal{M}^{(1)}$, since the lines of these matrices are not linearly independent. It can also be concluded that $\mathcal{M}^{(\ell)}$ $\ell = 0,1$ has one vanishing eigenvalue whose associated eigenvector is $\propto (1,1,...,1)$. We also assume that no other eigenvalues are zero.

The inversibility of the linear systems \eqref{eq:sys-0-mode} can be recovered by complementing them with the constraints imposed by the matching conditions. Indeed, from Eqs.~\eqref{eq:matching-kin} and \eqref{eq:phi-part-sol} we have, respectively,
\begin{subequations}
\label{eq:match-Phis}
\begin{align}
&
\label{eq:match-Phis-l=0}
\sum_{j} \beta J_{3,0}^{(j)} \Phi_{j,0} 
+
\sum_{j}
\left(-\left\langle \tau_{R,j}E_{{\bf p},j}  \mathcal{A}_{{\bf p},j} \right\rangle_{\widetilde{\mathrm{eq}},j}
+
\frac{1}{A^{(0)}_{j,0}} \left\langle  \mathcal{A}_{{\bf p},j} P^{(0)}_{{\bf p},j,0} \right\rangle_{\widetilde{\mathrm{eq}},j}  \beta J_{3,0}^{(j)} \right) \theta
\equiv 0,   
\\
&
\label{eq:match-Phis-l=1}
\sum_{j} J_{3,1}^{(j)} \Phi_{j,0}^{\mu} \equiv 0 .
\end{align}    
\end{subequations}
In Appendix \ref{sec:zero modes and m}, the procedure carried out for this inversion is discussed in detail. At the end, we obtain 
\begin{equation}
\label{eq:phi-0-sols-unif}
\begin{aligned}
&
\Phi_{i,0}
=
-
 \frac{\theta}{A_{i,0}^{(0)}}
 \langle \mathcal{A}_{{\bf p},i} P_{{\bf p},i,0}^{(0)} \rangle_{\widetilde{\mathrm{eq}},i}
+
\frac{\theta}{\beta \mathcal{J}_{3,0}}
 \sum_{k} \left\langle  \tau_{R {\bf p},k} E_{{\bf p},k}  \mathcal{A}_{{\bf p},k}  \right\rangle_{\widetilde{\mathrm{eq}},k},
\quad i = 1, \cdots, N_{\rm spec},
\\
&
\Phi_{i,0}^{\mu} = 0, \quad i = 1, \cdots, N_{\rm spec}.
\end{aligned}    
\end{equation}
We note that the first term of the equation for $\Phi_{i,0}$ above exactly cancels the second term in brackets in Eq.~\eqref{eq:phi-all-sol-scalar} for 
$\left. \Hat{\phi}^{(1)}_{{\bf p},i} \right\vert_{\ell = 0}$. \textcolor{black}{Besides, we remark that, in comparison to the single-particle case \cite{Rocha:2021zcw,Rocha:2022fqz}, determining the zero-mode coefficients $\Phi_{i,0}$ is more involved due to the fact that matching conditions and the new-RTA collision matrix contain sums over all particle species. In the single-particle case, a simple algebraic manipulation leads to the solution.}

\subsection{The first-order Chapman-Enskog solution}
\label{sec:1st-order solution}

Collecting the results of Eqs.~\eqref{eq:phi-part-sol} and \eqref{eq:phi-0-sols-unif}, we have the following result for the first order Chapman-Enskog contribution for the deviation function  
\begin{equation}
\label{eq:del-f-sol-1}
\begin{aligned}
&
\phi_{{\bf p},i}^{(1)} 
= 
\frac{f^{(1)}_{{\bf p},i}}{f_{0 {\bf p},i} \Tilde{f}_{0 {\bf p},i}}
=
F^{(0)}_{{\bf p},i} \theta +  F^{(2)}_{{\bf p},i} p^{\langle \mu}_{i}  p^{\nu \rangle}_{i} \sigma_{\mu \nu},
\end{aligned}    
\end{equation}
where we defined the scalar functions of momentum $F^{(0)}_{{\bf p},i}$ and $F^{(2)}_{{\bf p},i}$, 
\begin{subequations}
\label{eq:fin-del-f-comps}
\begin{align}
&
\label{eq:fin-del-f-comps-l=0a}
F^{(0)}_{{\bf p},i}
= 
- \frac{\tau_{R {\bf p},i}}{E_{{\bf p},i}} \mathcal{A}_{{\bf p},i}
+ 
\frac{E_{{\bf p},i}}{\mathcal{J}_{3,0}} \sum_{k} 
\left\langle \tau_{R{\bf p},k}E_{{\bf p},k} \mathcal{A}_{{\bf p},k} \right\rangle_{\widetilde{\mathrm{eq}},k} , 
\\
&
\label{eq:fin-del-f-comps-l=2}
F^{(2)}_{{\bf p},i} = \beta \frac{\tau_{R{\bf p},i}}{E_{{\bf p},i}},
\end{align}    
\end{subequations}
We note that the expression derived above work for arbitrary parametrizations of $\tau_{R{\bf p},i}$ in terms of momentum and of the particle index, $i$.

We remark that expression \eqref{eq:fin-del-f-comps-l=2} is identical to the corresponding result obtained assuming the Anderson-Witting RTA. On the other hand, the second term in expression \eqref{eq:fin-del-f-comps-l=0a} does not exist in the solution obtained with the Anderson-Witting RTA, emerging due to the counter-terms included in the novel RTA Ansatz discussed here. This new term must vanish when the relaxation time does not depend on momentum and particle species, given that we are already considering Landau matching conditions\footnote{For alternative matching conditions, the expression for $F^{(0)}_{{\bf p},i}$  will change \cite{Rocha:2021zcw,Rocha:2023ths}}. For instance, considering that $\tau_{R{\bf p},i} = t_{R} \ \forall i = 1, \cdots, N_{\rm spec}$, one can show that,
\begin{equation}
\begin{aligned}
&
\sum_{j} \left\langle \tau_{R,j}E_{{\bf p},j} \mathcal{A}_{{\bf p},j} \right\rangle
=
 t_{R} \sum_{j} \left\langle E_{{\bf p},j} \mathcal{A}_{{\bf p},j} \right\rangle
=
0,
\end{aligned}    
\end{equation}
where in the last step we have employed property \eqref{eq:AE-zero}. 

\textcolor{black}{Finally, it is important to remark some qualitative differences to the result obtained for a single-component gas. The correction $F^{(2)}_{{\bf p},i}$, that emerges due to the shear tensor, depends solely on the corresponding particle species and is identical to what is obtained in the single-component case. The correction $F^{(0)}_{{\bf p},i}$, on the other hand, is intrinsically different to the single-component case as the counter-terms introduced to restore energy-momentum conservation lead to a dependence on all particle species. Thus, the bulk viscous corrections will be more sensitive to the chemical composition of the gas.}

\subsection{Transport coefficients}

From Eq.~\eqref{eq:del-f-sol-1} and the definitions of the dissipative currents \eqref{eq:def-currents-kin}, we obtain the constitutive relations characteristic of Navier-Stokes theory,
\begin{subequations}
\label{eq:def-coeffs}
\begin{align}
\Pi & \simeq - \frac{1}{3} \sum_{j} \left\langle \left(\Delta^{\mu \nu} p_{j\mu} p_{j\nu}\right) \phi^{(1)}_{{\bf p},j} \right\rangle_{\widetilde{\mathrm{eq}},j}  \equiv - \zeta \theta  \\
\pi^{\mu \nu} & \simeq \sum_{j} \left\langle p^{\langle \mu}_{j} p^{ \nu \rangle}_{j} \phi^{(1)}_{{\bf p},j} \right\rangle_{\widetilde{\mathrm{eq}},j} \equiv  2 \eta \sigma^{\mu \nu},
\end{align}
\end{subequations}
where we identify $\zeta$ as the bulk viscosity and $\eta$ as the shear viscosity. The microscopic expressions for these transport coefficients are 
\begin{subequations}
\label{eq:1st-ord-expr-no-ch}
\begin{align}
\zeta &= -\frac{1}{3}\sum_{i}
\left\langle \left(\Delta^{\mu \nu} p_{i\mu} p_{i\nu}\right) 
\frac{\tau_{R{\bf p},i}}{E_{{\bf p},i}} \mathcal{A}_{{\bf p},i} \right\rangle_{\widetilde{\mathrm{eq}},i}
-
\frac{\mathcal{J}_{3,1}}{\mathcal{J}_{3,0}}\sum_{k} \left\langle \tau_{R{\bf p},k}E_{{\bf p},k} \mathcal{A}_{{\bf p},k} \right\rangle_{\widetilde{\mathrm{eq}},k} 
\notag
\\
&
=
\sum_{i}
\left\langle  
\frac{\tau_{R{\bf p},i}}{E_{{\bf p},i}} \left[ \mathcal{A}_{{\bf p},i} \right]^{2} \right\rangle_{\widetilde{\mathrm{eq}},i},
\label{eq:1st-ord-expr-no-ch-zeta}
\\
\label{eq:1st-ord-expr-no-ch-eta}
\eta & = \frac{\beta}{15} \sum_{i} \left\langle \left(\Delta^{\mu \nu} p_{i\mu} p_{i\nu}\right)^{2}\frac{\tau_{R{\bf p},i}}{E_{{\bf p},i}} \right\rangle_{\widetilde{\mathrm{eq}},i},
\end{align}    
\end{subequations}
where, from the first to the second line of Eq.~\eqref{eq:1st-ord-expr-no-ch-zeta}, we have employed the definition of the function $\mathcal{A}_{{\bf p},i}$ (see Eq.~\eqref{eq:chap-ensk-1st-ord-b}). It is noted that, in general, the expression for the bulk viscosity coefficient depends on the matching conditions employed \cite{Rocha:2022fqz,Rocha:2023ths} and the expression in the second line is only valid for Landau matching conditions. \textcolor{black}{ We note that that the above microscopic formulas for the transport coefficients are qualitatively similar to the matching-invariant coefficients, $\zeta_{s}$ and $\eta_{s}$, displayed in Eq.~(63) of Ref.~\cite{Rocha:2022fqz}, for a single-component gas of quasi-particles -- the crucial difference is that here each expression is summed over all particle species}. 
In the next section, we shall relate these expressions for the transport coefficients to the entropy production.
We also note that the expressions above admit an arbitrary parametrization of the relaxation times, $\tau_{R{\bf p},i}$, with the momenta of the particles. In Sec.~\ref{sec:results}, we shall analyze the behavior of the transport coefficients for a specific parametrization of the relaxation times and consider the species in hadron-resonance gas model. 

\section{Entropy production}
\label{sec:entropy-production}

One of the main features of the Boltzmann equation is its compatibility with the second law of thermodynamics. This connection is made through the $H$-theorem \cite{deGroot:80relativistic,cercignani:02relativistic}. Summarily, it states that there exists a vector functional $S^{\mu}$ of $f_{\bf p}$ such that $\partial_{\mu}S^{\mu} \geq 0$. Then, from the connection with thermodynamics, one identifies this functional with the entropy 4-current and use the particular instance $\partial_{\mu}S^{\mu} = 0$ as the definition of the equilibrium state\footnote{The molecular chaos hypothesis, which states that the particles are uncorrelated prior to the collision processes, is responsible for breaking  time-reversal and thus rendering the Boltzmann equation compatible with the second law of thermodynamics.}. Hence, in addition to properties \eqref{eq:lin-coll-csv-law}, \eqref{eq:self-adj}, and \eqref{eq:zero-modes-L}, a physically consistent approximation to the collision integral should give rise to a non-negative entropy production. In kinetic theory, the entropy 4-current is defined by \cite{deGroot:80relativistic} 
\begin{equation}
\label{eq:s-mu-FD-BE}
\begin{aligned}
S^{\mu} = - \sum_{i} \int dP_{i}\, p^{\mu} \left[ \frac{\Tilde{f}_{{\bf p},i}}{a_{i}} \ln \Tilde{f}_{{\bf p},i}  + f_{{\bf p},i} \ln f_{{\bf p},i} \right].
\end{aligned}
\end{equation}
Then, taking the Boltzmann equation into account, the four-divergence of $S^{\mu}$ can be computed to be 
\begin{equation}
\label{eq:entropy-prod-C}
\begin{aligned}
\partial_{\mu} S^{\mu} = - \sum_{i} \int dP\, C_{i}[f_{\mathbf{p}}] ( \ln f_{\mathbf{p},i} - \ln \Tilde{f}_{\mathbf{p},i} ),
\end{aligned}
\end{equation}
hence the collision terms possesses a prominent role in the entropy production analysis. In fact, if $C_{i}[f_{\mathbf{p}}] = 0$, $\forall i = 1, \cdots, N_{\rm spec}$, the entropy production is zero and the system is in local equilibrium.  

Considering that the system of particles is sufficiently close to equilibrium it is sensible to linearize the above expression as 
\begin{equation}
\label{eq:entropy-prod-C-ln}
\begin{aligned}
\partial_{\mu} S^{\mu} = - \sum_{i} \int dP\, C_{i}[f_{\mathbf{p}}] ( \ln f_{\mathbf{p},i} - \ln \Tilde{f}_{\mathbf{p},i} ) 
\simeq
- \sum_{i} \int dP_{i}\, f_{0\mathbf{p},i} \phi_{\mathbf{p},i} \hat{L}\phi_{\mathbf{p},i}
,
\end{aligned}
\end{equation}
where it was used that $f_{\bf p} = f_{0{\bf p}}(1 + \Tilde{f}_{0{\bf p}} \phi_{\bf p})$, $\Tilde{f}_{\bf p} = \Tilde{f}_{0{\bf p}}(1 - a f_{0{\bf p}} \phi_{\bf p})$ and the approximation $\ln(1+x) \simeq x$. Within the novel RTA, Eq.~\eqref{eq:NRTA}, we can derive the following expression, employing the expansion \eqref{eq:expn-tensorial}
\begin{equation}
\label{eq:smu-prod-Phis}
\begin{aligned}
\partial_{\mu} S^{\mu} & = \sum_{i,j} \mathcal{M}^{(0)}_{ij} \Phi_{i,0} \Phi_{j,0} 
+
\frac{1}{3}\sum_{i,j} \mathcal{M}^{(1)}_{ij} \Phi_{i,0}^{\mu}\Phi_{j,0 \mu } 
+
\sum_{i} \left(\sum_{n=1}^{\infty} A^{(0)}_{i,n} \Phi_{i,n}^{2} 
+
\frac{1}{3}\sum_{n=1}^{\infty} A^{(1)}_{i,n} \Phi_{i,n}^{\mu}\Phi_{i,n \mu} \right. \\ 
& \left.
+
\sum_{\ell = 2}^{\infty}\frac{\ell!}{(2\ell + 1)!!}\sum_{n=0}^{\infty} A^{(\ell)}_{i,n} \Phi_{i,n}^{\mu_{1} \cdots \mu_{\ell}}\Phi_{i,n \mu_{1} \cdots \mu_{\ell}} \right), 
\end{aligned}
\end{equation}
which is composed of a sum of quadratic terms, which are manisfestly non-negative, since, from definition \eqref{eq:orth-polyn-ell}, $(-1)^{\ell} A_{i,n}^{(\ell)} \geq 0$ and we also have that $(-1)^{\ell} \Phi_{i,n}^{\mu_{1} \cdots \mu_{\ell}}\Phi_{i,n \mu_{1} \cdots \mu_{\ell}} \geq 0$. The only terms that are not manifestly non-negative so far comprise scalar and vector components corresponding to zero modes of the linearized collision term, $\Phi_{i,0}$ and $\Phi_{i,0}^{\mu}$. Their non-negativeness can be derived using the Cauchy-Schwarz inequality,
\begin{equation}
\label{eq:cauchy-schw}
 \left(\sum_{i} a_{i} b_{i} \right)^{2} \leq \left(\sum_{i} a^{2}_{i} \right) \left(\sum_{i} b^{2}_{i} \right),   
\end{equation}
which is valid for arbitrary $a_{i}$ and $b_{i}$. Indeed, the remaining scalar and vector terms in Eq.~\eqref{eq:smu-prod-Phis} can be expressed as
\begin{equation}
\begin{aligned}
\sum_{i,j} \mathcal{M}^{(0)}_{ij} \Phi_{i,0} \Phi_{j,0} & = 
\sum_{i} A^{(0)}_{i,0}  \Phi_{i,0}^{2}
-
 \frac{\left(\sum_{i} A^{(0)}_{i,0} \Phi_{i,0} \right)^{2 }}{\sum_{i} A^{(0)}_{i,0} }   \\
\sum_{i,j} \mathcal{M}^{(1)}_{ij} \Phi_{i,0}^{ \mu }\Phi_{j,0 \langle \mu \rangle}
&=
\sum_{i} A_{i,0}^{(1)} \Phi_{i,0}^{ \mu }\Phi_{i,0 \mu }
-
\frac{\left(\sum_{i} A_{i,0}^{(1)} \Phi_{i,0}^{\mu}\right) \left(\sum_{i} A_{i,0}^{(1)} \Phi_{i,0 \mu}\right)}{\sum_{i} A_{i,0}^{(1)}},
\end{aligned}    
\end{equation}
whose non-negativeness can be derived from the inequalities \eqref{eq:cauchy-schw} by taking, $a_{i} = \sqrt{A_{i,0}^{(0)}} \Phi_{i,0}  $, $b_{i} = \sqrt{A_{i,0}^{(0)}}$ for the first expression above and, without loss of generality, in the local rest frame of the fluid (LRF), taking $a_{i} = \sqrt{-A_{i,0}^{(1)}} \Phi_{i,0}^{m}\vert_{\rm LRF}  $, $b_{i} = \sqrt{-A_{i,0}^{(1)}}$, where $\Phi_{i,0}^{m}\vert_{\rm LRF}$ denotes the space components of the corresponding vector, since the property $\Phi_{i,0}^{\mu}u_{\mu} = 0$ implies that $\Phi_{i,0}^{0} \vert_{\rm LRF} = 0$.  This establishes the consistency with the second law of thermodynamics for arbitrary values of $\Phi$-components within the novel RTA prescription. Employing the Chapman-Enskog expansion solution obtained in Sec.~\ref{sec:1st-order-CE}, one can express the entropy production as,
\begin{equation}
\label{eq:H-thm-smu}
\begin{aligned}
& \partial_{\mu} S^{\mu} = \beta \zeta \theta^{2} + 2 \eta \beta \sigma^{\mu \nu} \sigma_{\mu \nu},
\end{aligned}    
\end{equation}
which is also manifestly non-negative since $\zeta \geq 0$ and $\eta \geq 0$. 

\section{Non-equilibrium deviations for the Hadron-resonance gas}
\label{sec:results}

The Hadron-Resonance Gas model (HRG) \cite{Venugopalan:1992hy,Karsch:2003vd,Huovinen:2009yb,Bazavov:2009zn,Borsanyi:2010cj,Ratti:2021ubw,Vogt:2007zz} is suitable for the description of nuclear matter at low temperatures. In simulations of heavy-ion collisions, this weakly-interacting system is employed in the description of the late stages, where matter has sufficiently cooled down to be described in terms of hadronic degrees of freedom instead of deconfined quarks and gluons. In such change of description, multistage codes of heavy-ion collisions often employ the Chapman-Enskog formalism to calculate the non-equilibrium corrections to the equilibrium distribution for each hadronic species sampled. 

In the following, we apply the formalism constructed in Secs.~\ref{sec:nRTA-mult-part} and \ref{sec:1st-order-CE} to a hadron resonance gas. We then calculate the deviation function for each particle species, $\phi_{{\bf p},i}$, with $i = 1, \cdots, N_{\rm spec}$ and $N_{\rm spec}$ denoting the total number of hadronic species considered, following the Chapman-Enskog procedure. We shall also employ the following parametrization of the relaxation time 
\begin{equation}
\label{eq:polyn-dep-tR}
\tau_{R{\bf p},i} = t_{R} \left( \frac{E_{{\bf p},i}}{T} \right)^{\gamma}, 
\end{equation} 
where $t_{R}$ is a timescale that depends neither on the momentum of a given particle species nor on the particle species index $i$, but can be a function of temperature. Besides, $\gamma$ is a phenomenological parameter (that will be assumed to be the same for all particle species) and encompasses the microscopic interaction between them. For instance, it was previously demonstrated that $\gamma = 1$ emulates the kinetic regime of scalar fields self-interacting with a quartic potential \cite{Calzetta:1986cq,Denicol:2022bsq}. 

In particlization models \cite{McNelis:2019auj,Shen:2014vra}, in which particles are sampled from fluid cells, the results for the deviation functions, $\phi_{{\bf p},i}$, are more conveniently expressed in terms of the dissipative currents, $\Pi$ and  $\pi^{\mu \nu}$, instead of the gradients $\theta$ and $\sigma^{\mu \nu}$. Thus, we use Eqs.~\eqref{eq:del-f-sol-1} and the constitutive relations \eqref{eq:def-coeffs}, to write
\begin{subequations}
\label{eq:CF1}
\begin{align}
\label{eq:CF1-p}
\phi_{{\bf p},i}^{(1)} 
& =
\left. \phi_{{\bf p},i}^{(1)} \right\vert_{\mathrm{bulk}}
+
\left. \phi_{{\bf p},i}^{(1)} \right\vert_{\mathrm{shear}}  ,
\\
\label{eq:CF1-bulk}
\left. \phi_{{\bf p},i}^{(1)} \right\vert_{\mathrm{bulk}}
&=
- \frac{F^{(0)}_{{\bf p},i}}{\zeta} \Pi,
\\
\label{eq:CF1-shear}
\left. \phi_{{\bf p},i}^{(1)} \right\vert_{\mathrm{shear}}  
&=
\frac{F^{(2)}_{{\bf p},i}}{2 \eta} p^{\langle \mu}_{i}  p^{\nu \rangle}_{i} \pi_{\mu \nu},
\end{align}    
\end{subequations}
where in the latter couple of equations, we have defined, the bulk and shear sectors of the deviation function, $\left. \phi_{{\bf p},i}^{(1)} \right\vert_{\mathrm{bulk}}$ and $\left. \phi_{{\bf p},i}^{(1)} \right\vert_{\mathrm{shear}}$, respectively. In what follows we shall discuss each sector in more detail.

\subsection{Bulk sector}
\label{sec:bulk-sector-non-eq}

As it was seen in Sec.~\ref{sec:1st-order solution}, the expressions for the non-equilibrium corrections to the single particle distribution function are more involved due to the matching conditions. Substituting the parametrization \eqref{eq:polyn-dep-tR} into the results derived in Eqs.~\eqref{eq:fin-del-f-comps-l=0a}, \eqref{eq:1st-ord-expr-no-ch-zeta}, we have
\begin{equation}
\label{eq:phi_bulk}
\begin{aligned}
&
\left. \phi_{{\bf p},i}^{(1)} \right\vert_{\mathrm{bulk}}  
=
\left[
\varphi(T) \frac{E_{{\bf p},i}}{T}
- 
\frac{m_{i}^{2}}{3 T^{2}} \left( \frac{E_{{\bf p},i}}{T} \right)^{\gamma - 1}  
+ 
\left( \frac{1}{3} - c_{s}^{2} \right) \left( \frac{E_{{\bf p},i}}{T} \right)^{\gamma + 1} 
 \right] \frac{\Pi}{B_{\Pi}},
\end{aligned}
\end{equation}
where we defined the coefficients
\begin{subequations}
\begin{align}
\label{eq:varphi}
\varphi(T)
&=
\frac{1}{T^{\gamma}} \frac{1}{\mathcal{J}_{3,0}} \left(c_{s}^{2} \mathcal{J}_{\gamma+3,0} 
-  
\mathcal{J}_{\gamma+3,1} \right),
\\
\label{eq:B_PI}
B_{\Pi} & = \frac{\zeta}{t_{R}}
=
\frac{1}{T^{\gamma+1}}
\left(- c_{s}^{2} \mathcal{J}_{\gamma+3,1} + \frac{5}{3} \mathcal{J}_{\gamma+3,2} \right) 
+
\frac{1}{T^{\gamma+1}}\frac{\mathcal{J}_{3,1}}{\mathcal{J}_{3,0}}
\left(c_{s}^{2} \mathcal{J}_{\gamma+3,0} 
-  
\mathcal{J}_{\gamma+3,1} \right),
\end{align}
\end{subequations}
and we note that the parametrization \eqref{eq:polyn-dep-tR} determines the energy dependence of the correction $\left. \phi_{{\bf p},i}^{(1)} \right\vert_{\mathrm{bulk}}$. In contrast to the result obtained with the usual Anderson-Witting RTA (which violates energy-momentum conservation when $\gamma \neq 0$), we obtain a new contribution proportional to $E_{{\bf p},i}/T$. We note that, unless $\gamma=0$, the relaxation time coefficient $t_{R}$ does not coincide with the bulk viscosity relaxation time \cite{Rocha:2021lze}.

Now we assess the behavior of the bulk sector coefficients in the high temperature regime, where $a_{i} \to 0$, $m_{i}/T \to 0$, $i = 1, \cdots, N_{\rm spec}$. For the coefficient $\varphi(T)$, we have
\begin{equation}
\begin{aligned}
&
\varphi(T) \approx \frac{\Gamma(\gamma + 3)}{72} \left( 1 - \frac{(\gamma + 3)(\gamma + 4)}{12} \right) \frac{\sum_{i} g_{i}\left( m_{i}/T \right)^{2}}{\sum_{i} g_{i}},
\end{aligned}    
\end{equation}
where $\Gamma(s)$ denotes the Euler gamma function \cite{NIST:DLMF}. The above expression shows that this coefficient vanishes if the masses of the particles are set to zero, with the leading order in mass dependence being $\mathcal{O}((m/T)^{2})$. This is similar to the behavior of the conformal violation of the speed of sound, which also contributes to the non-equilibrium correction of Eq.~\eqref{eq:phi_bulk},
\begin{equation}
\begin{aligned}
&
\frac{1}{3} - c_{s}^{2}
\approx
\frac{1}{36}
\frac{\sum_{i} g_{i}\left( m_{i}/T \right)^{2}}{\sum_{i} g_{i}}.
\end{aligned}    
\end{equation}
Moreover, since the parameter $\gamma$ appears in the argument of the $\Gamma$-function, it is expected that the range of values of $\varphi(T)$ changes drastically as $\gamma$ changes. This feature will be observed for all of the remaining coefficients. In this regime, the normalized bulk viscosity coefficient reads  
\begin{equation}
\begin{aligned}
&
\frac{B_{\Pi}}{\varepsilon_{0} + P_{0}} 
\approx 
 \frac{\Gamma(\gamma + 1)}{72} \frac{\sum_{i} g_{i}\left( m_{i}/T \right)^{4}}{\sum_{i} g_{i}} - 
\frac{\Gamma(\gamma + 3)}{432} \left( 1 - \frac{(\gamma + 3)(\gamma + 4)}{24} \right) \left[\frac{\sum_{i} g_{i}\left( m_{i}/T \right)^{2}}{\sum_{i} g_{i}} \right]^{2},
\end{aligned}   
\end{equation}
which possess an effective $\mathcal{O}((m/T)^{4})$ at leading order in the masses of the particles. Thus, it vanishes faster than $\varphi(T)$ as the mass-to-temperature ratios are taken to zero. In fact, $B_{\Pi}/(\varepsilon_{0} + P_{0}) \propto (1/3 - c_{s}^{2})^{2}$ in this regime. We also note that when $\gamma \to 0$, our present results reduce to the one in Eq.~(25a) of Ref.~\cite{Rocha:2024rce}, where the method of moments was employed to derive transient hydrodynamic equations of motion.

In the opposite limit, at low temperatures, where $m_{i}/T \to \infty$, $i = 1, \cdots, N_{\rm spec}$, the bulk sector coefficients behave as 
\begin{subequations}
\begin{align}
\varphi(T) & \approx \frac{1}{\mathfrak{m}_{7/2}} \left[\frac{\mathfrak{m}_{5/2}}{\mathfrak{m}_{7/2}}
\mathfrak{m}_{\gamma + 7/2}
-
\mathfrak{m}_{\gamma + 5/2}
\right],
\\
\label{eq:low-T-bulk}
\frac{B_{\Pi}}{\varepsilon_{0} + P_{0}} 
&\approx 
\frac{1}{\mathfrak{m}_{5/2}}
\left[
\frac{5}{3}
\mathfrak{m}_{\gamma + 3/2}
-
2 \frac{\mathfrak{m}_{5/2}}{\mathfrak{m}_{7/2}}
\mathfrak{m}_{\gamma + 5/2}
+
\left(\frac{\mathfrak{m}_{5/2}}{\mathfrak{m}_{7/2}} \right)^{2}
\mathfrak{m}_{\gamma + 7/2}
\right],
\\
\label{eq:low-T-bulk-notation}
\mathfrak{m}_{A} & \equiv \sum_{i} g_{i} (m_{i}/T)^{A} e^{-m_{i}/T}.
\end{align}   
\end{subequations}
Now we observe that both of the coefficients depend on weighted sums of powers of the particle masses and that the power of the leading order contribution for both coefficients depends on $\gamma$. Indeed, both coefficients behave asymptotically as $\mathcal{O}((m/T)^{\gamma - 1})$.

\begin{figure}[!h]
    \centering
\begin{subfigure}{0.5\textwidth}
    \includegraphics[scale=0.35]{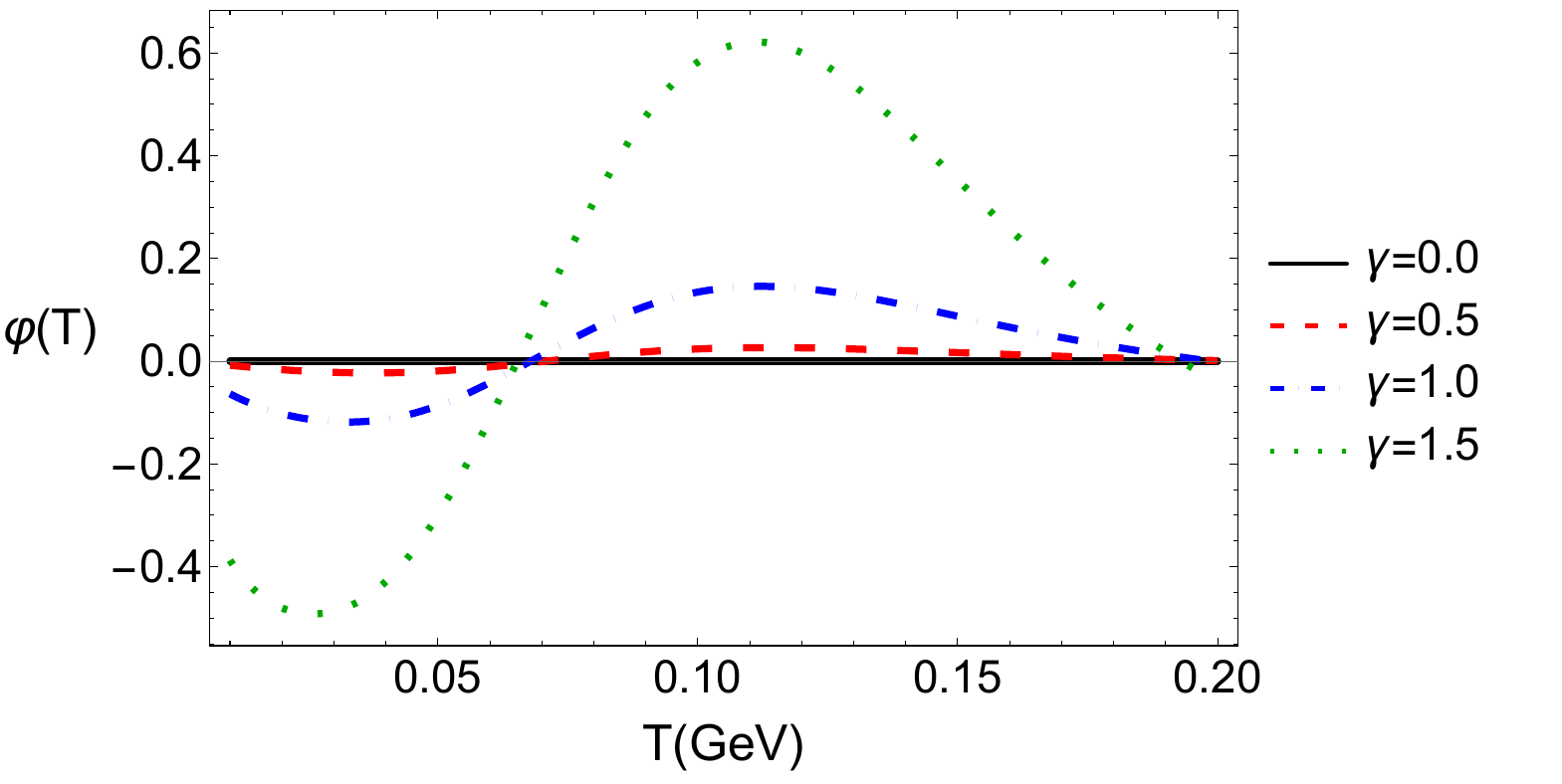}
    \caption{Coefficient $\varphi(T)$ (cf.~Eq.~\eqref{eq:varphi})}
    \label{fig:varphi}    
\end{subfigure}\hfil    
\begin{subfigure}{0.5\textwidth}
    \includegraphics[scale=0.38]{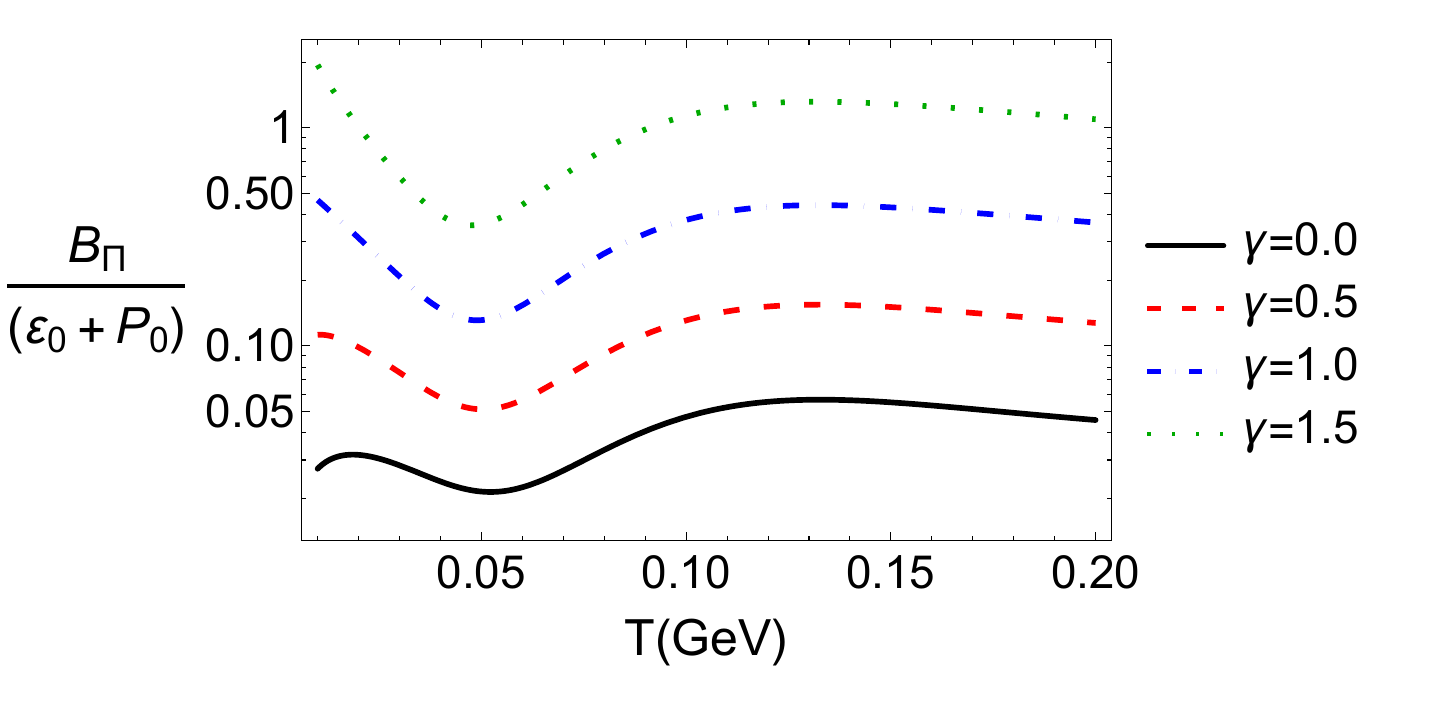}
    \caption{Normalized $B_{\Pi} = \zeta/t_{R}$ (cf.~Eq.~\eqref{eq:B_PI}) coefficient }
    \label{fig:beta-bulk}    
\end{subfigure}\hfil
\begin{subfigure}{0.5\textwidth}
    \includegraphics[scale=0.33]{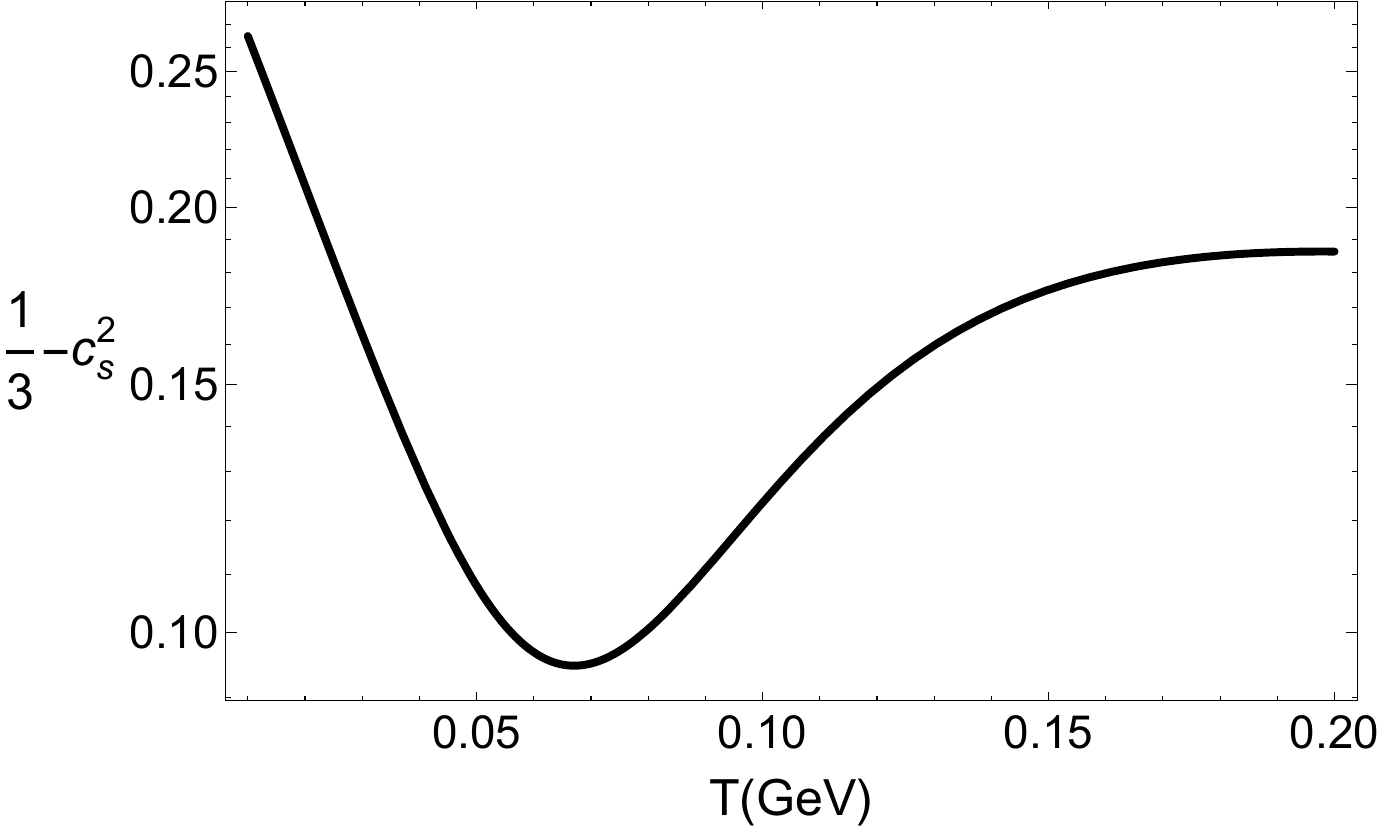}
    \caption{Conformal violation of the speed of sound}
    \label{fig:1/3-cs2}    
\end{subfigure}\hfil
\begin{subfigure}{0.5\textwidth}
    \includegraphics[scale=0.35]{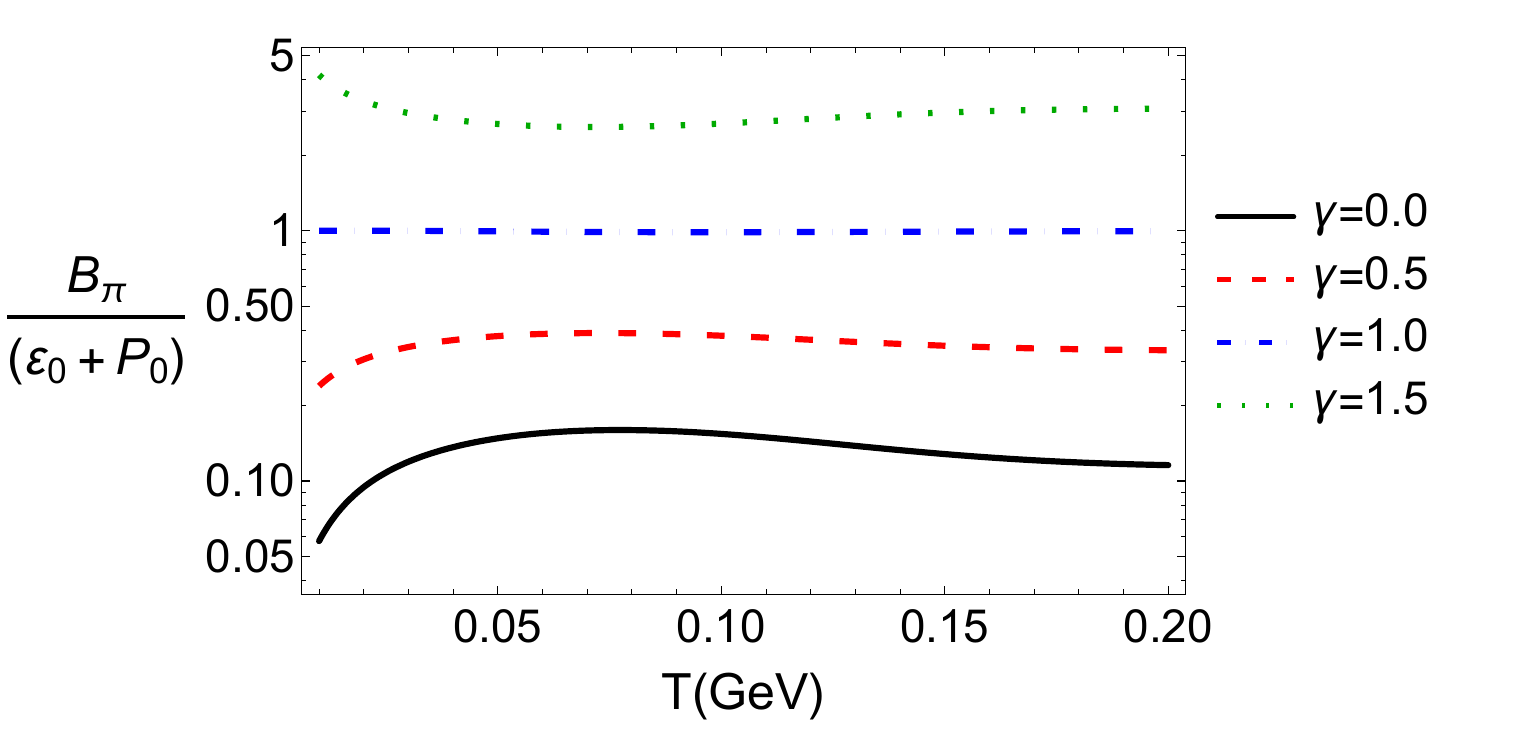}
    \caption{Normalized $B_{\pi} = \eta/t_{R}$ coefficient (cf.~Eq.~\eqref{eq:B_shear})}
    \label{fig:beta-shear}    
\end{subfigure}\hfil
\caption{Normalized coefficients as a function of temperature for the hadron-resonance gas computed numerically}
\label{fig:1st-order-coeffs}
\end{figure}

In Fig.~\ref{fig:varphi}, we display the coefficient $\varphi(T)$ as a function of temperature. In this figure and the remaining ones, the UrQMD \cite{Bass:1998ca,Bleicher:1999xi} particle list was employed. We see that, in the range of temperatures displayed, this coefficient has a non-monotonic behavior, being negative at $T \approx 0.01$ GeV and then increasing in value until it becomes positive around $T \approx 0.07$ GeV, reaching a maximum value at $T \approx 0.11$ GeV and decreasing in magnitude once again. The range of values in which this ``oscillation'' occurs grows significantly as $\gamma$ grows. For $\gamma = 1.0$, this coefficient is comparable in magnitude to $(1/3 -c_{s}^{2})$, displayed in Fig.~\ref{fig:1/3-cs2}. In Fig.~\ref{fig:beta-bulk}, we show the behavior of the normalized bulk viscosity coefficient $B_{\Pi}/(\varepsilon_{0} + P_{0})$ as a function of temperature. Clearly, it displays a non-monotonic temperature behavior which depends significantly on the parameter $\gamma$. The asymptotic behavior at small temperatures is consistent with the $\mathcal{O}((m/T)^{\gamma - 1})$ behavior found in Eq.~\eqref{eq:low-T-bulk}.

\subsection{Shear sector}

Collecting the results in Eqs.~\eqref{eq:del-f-sol-1},  \eqref{eq:fin-del-f-comps-l=2}, \eqref{eq:1st-ord-expr-no-ch-eta}, and parametrization \eqref{eq:polyn-dep-tR}, we have the following result for the shear sector of non-equilibrium deviations 
\begin{subequations}
\label{eq:phi_shear}
\begin{align}
\left. \phi_{{\bf p},i}^{(1)} \right\vert_{\mathrm{shear}}  
& =
\frac{1}{T^{2}} \left( \frac{E_{{\bf p},i}}{T} \right)^{\gamma-1}   
 p^{\langle \mu}_{i}  p^{\nu \rangle}_{i} \frac{\pi_{\mu \nu}}{2 B_{\pi}},
 \\
\label{eq:B_shear} 
B_{\pi} & = \frac{\eta}{t_{R}}
=
\frac{1}{T^{\gamma+1}}\mathcal{J}_{\gamma+3,2},
\end{align}        
\end{subequations}
where we see that the parametrization \eqref{eq:polyn-dep-tR} affects the dependence of the correction $\left. \phi_{{\bf p},i}^{(1)} \right\vert_{\mathrm{shear}}$ on the energy of the particles and also the coefficient $B_{\pi}$, since both depend on the parameter $\gamma$. In contrast to the case of bulk viscosity contributions, we note that our prescription \eqref{eq:NRTA} reduces to that of Anderson and Witting for all values of $\gamma$. We note that our results also reduce to the ones in Refs.~\cite{Rocha:2022fqz,Rocha:2024rce}. 

In the high temperature limit, where $m_{i}/T \to 0$, $i = 1, \cdots, N_{\rm spec}$, the normalized shear viscosity coefficient reduces, at leading order, to
\begin{equation}
\label{eq:high-T-shear}
\begin{aligned}
&
\frac{B_{\pi}}{\varepsilon_{0} + P_{0}} 
\approx
\frac{\Gamma(\gamma + 5)}{120}. 
\end{aligned}   
\end{equation}
We see that, in this limit, the normalized shear coefficient depends only on the phenomenological parameter $\gamma$. For $\gamma = 0$, the above expression reduces to $B_{\pi}/(\varepsilon_{0} + P_{0}) \approx 1/5$. On the other hand, for $\gamma = 1$, we have $B_{\pi}/(\varepsilon_{0} + P_{0}) \approx 1$.

Also, in the low temperature limit, where $m_{i}/T \to \infty$, $i = 1, \cdots, N_{\rm spec}$, we have
\begin{equation}
\label{eq:low-T-shear}
\begin{aligned}
&
\frac{B_{\pi}}{\varepsilon_{0} + P_{0}} \approx \frac{\mathfrak{m}_{\gamma + 3/2}}{\mathfrak{m}_{5/2}},
\end{aligned}   
\end{equation}
where we employed the notation defined in Eq.~\eqref{eq:low-T-bulk-notation} for the weighted averages of powers of the masses of the different particle species. In general, we note that the effective $m/T$ power of this contribution is $B_{\pi}/(\varepsilon_{0} + P_{0}) \approx \mathcal{O}((m/T)^{\gamma -1})$. Indeed, for $\gamma = 1$, $B_{\pi}/(\varepsilon_{0} + P_{0}) \approx 1$. 

In Fig.~\ref{fig:beta-shear}, we show the behavior of the normalized coefficient $B_{\pi}/(\varepsilon_{0} + P_{0})$ as a function of temperature. Then, we see, as in Refs.~\cite{Rocha:2021zcw,Rocha:2022fqz}, the range of values reached by this transport coefficient changes drastically as $\gamma$ varies. It is also seen that at low temperatures, for values of $\gamma < 1$ the coefficient $B_{\pi}/(\varepsilon_{0} + P_{0})$ decreases, whereas for $\gamma > 1$ this coefficient increases. For $\gamma = 1$, the normalized coefficient remains practically constant for all temperatures, $B_{\pi}/(\varepsilon_{0} + P_{0}) \approx 1$. This stems from the identity\footnote{This identity is valid for all temperatures and, in general, $\mathcal{J}_{n,q} =T [\mathcal{I}_{n-1,q-1} + (n-2q)\mathcal{I}_{n-1,q} ]$} $\mathcal{J}_{4,2} = T \mathcal{I}_{3,1}$ (see Eq.~\eqref{eq:thermodynamic_integrals}) and the fact that $\mathcal{I}_{3,1} \approx \mathcal{J}_{3,1} = T (\varepsilon_{0} + P_{0})$ in the range of temperature displayed.  We also note that this behavior is consistent with the $\mathcal{O}((m/T)^{\gamma - 1})$ behavior of Eqs.~\eqref{eq:high-T-shear} and \eqref{eq:low-T-shear}.

\section{Conclusion}
\label{sec:conclusion}

In the present paper, we have presented the multi-particle counterpart of the RTA ansatz proposed in Ref.~\cite{Rocha:2021zcw} for systems where only the conservation of energy and momentum are relevant. We improve upon the Anderson-Witting RTA Ansatz by adding counter-terms which are related to microscopically-conserved quantities. In the present case, they are the components of the four-momentum, $p^{\mu}_{i}$, for each species $i = 1, \cdots, N_{\rm spec}$. Our prescription recovers fundamental properties of the linearized collision term, which are essential to maintain the Boltzmann equation compatible with the local conservation laws. These properties and, by extension, the macroscopic conservation laws are not obeyed by the Anderson-Witting RTA ansatz when the relaxation time depends on the momentum of the particles or when Landau prescription to define hydrodynamic variables are not employed.

As an application, we have derived first-order Chapman-Enskog non-equilibrium corrections for the single-particle distribution functions for a gas consisting of particle species with arbitrary masses, $m_{i}$, $i = 1, \cdots, N_{\rm spec}$. This is performed for an arbitrary parametrization of the RTA timescale, $\tau_{R{\bf p},i}$, with the momentum of the particles and for Landau matching conditions (as long as the corresponding integrals do not diverge). At first order, this procedure requires the inversion of the collision term. This becomes non-trivial within our RTA prescription, since, by construction, the components of the four-momentum of the particles, $p^{\mu}_{i}$, are zero-modes of the collision term, which have to be removed from the inversion procedure, since they render the determinant of the collision term matrix to be zero. Nevertheless, an invertible system of linear equations can be derived taking the matching conditions into account. From the latter, constraints for the expansion coefficients in Eq.~\eqref{eq:expn-tensorial} are derived. The removal of zero-modes is implemented by explicitly substituting the expansion coefficient for one of the particles into the linear system formed by integrating the Boltzmann equation with the zero modes (see Eq.~\eqref{eq:subsys-num-eff-2}). After the corresponding system of equations is inverted, the non-equilibrium corrections are expressed in a manner that does not depend on the particle chosen in the zero-mode removal procedure. 

After that, employing parametrization \eqref{eq:polyn-dep-tR} for the relaxation time of each particle species, we derive expressions for the deviation functions for the hadron-resonance gas, which can be employed in particlization models. This parametrization introduces the parameter $\gamma$ which controls the momentum dependence of the RTA timescale, $\tau_{R{\bf p},i}$, for each particle species with its energy. In the limit $\gamma \to 0$, $\tau_{R{\bf p},i} = t_{R}$, $\forall i = 1, \cdots, N_{\rm spec}$, the RTA timescale does not depend on momentum and our results reduce to the the ones of the Anderson-Witting RTA. In this case, non-equilibrium corrections can be expressed in terms of polynomials in $(E_{{\bf p},i}/T)$ and of the  coefficients $B_{\Pi} = \zeta/t_{R}$  and $\varphi(T)$ for the bulk sector and $B_{\pi} = \eta/t_{R}$ for the shear sector. The coefficient $\varphi(T)$ is essentially new and controls the contribution of the $(E_{{\bf p},i}/T)$ to the non-equilibrium corrections.  In the limit where $\gamma \to 0$, these coefficients reduce respectively to $\beta_{\Pi} = \zeta/\tau_{\Pi}$ and $\beta_{\pi} = \eta/\tau_{\pi}$. However, this correspondence cannot be trivially made when $\gamma \neq 0$. Moreover, $\varphi(\gamma \to 0, T) = 0$. In the high temperature limit, $B_{\pi}$ becomes a constant that depends only on $\gamma$, whereas $\varphi(T)$ behaves similarly to $1/3 - c_{s}^{2}$ and $B_{\Pi}$ behaves as $(1/3 - c_{s}^{2})^{2}$. The latter expression reduces to the one in Eq.~(25a) of Ref.~\cite{Rocha:2024rce}. In the future, it would be of interest to assess the effects of these results to particle observables in simulation chains, to assess the role of baryon chemical potential \textcolor{black}{ and of alternative matching conditions. Furthermore, it would be important to verify how much the results of this paper are affected by certain details of the hadron resonance gas model, such as resonance widths, as originally investigated in \cite{Vovchenko:2019pjl}.}

\section*{Acknowledgements}

The authors thank T.~Nunes and I.~Aguiar for fruitful discussions. G.~S.~R. is funded by Vanderbilt University and was also funded in part by the U.S. Department of Energy, Office of Science under Award Number DE-SC-0024347, Coordenação de Aperfeiçoamento de Pessoal de Nível Superior (CAPES) Finance code 001, award No. 88881.650299/2021-01 and by Conselho Nacional de Desenvolvimento Científico e Tecnológico (CNPq), grant No.~142548/2019-7. G.S.D.~also acknowledges CNPq as well as Fundação Carlos Chagas Filho de Amparo à Pesquisa do Estado do Rio de Janeiro (FAPERJ), Grant No.~E-26/202.747/2018.

\appendix

\section{Series summation procedure}
\label{apn:sum-coeffs}

In this Appendix, we discuss the intermediate steps which involve summations of series in Eqs.~\eqref{eq:phi-all-sol}. To that end, we will see that it is not necessary to explicitly construct the set of polynomials, but only use their orthogonality, Eq.~\eqref{eq:orth-polyn-ell}. Therefore, we assume that the set of polynomials $P^{(\ell)}_{{\bf p},i, n}$ ($n=0,1,...$, for a given $\ell$ and a given particle species, $i$) is a complete basis of the space of functions of momentum with a finite norm, given by the left-hand side of Eq.~\eqref{eq:orth-polyn-ell}, so that an arbitrary function of energy can be expanded as 
\begin{equation}
\label{eq:expn-Pnl}
\begin{aligned}
& G(E_{\mathbf{p},i}) =  \sum_{n=0}^{\infty} \lambda_{i,n}^{(\ell)} P^{(\ell)}_{{\bf p},i, n}, 
\end{aligned}
\end{equation}
for a given $\ell$. To find the coefficients $\lambda_{i,n}^{(\ell)}$, one resorts to the orthogonality relation \eqref{eq:orth-polyn-ell}. Thus, integrating both sides with $(E_{\mathbf{p},i}/\tau_{R{\bf p},i}) \left( \Delta^{\mu \nu} p^{\mu}_{i} p^{\nu}_{i} \right)^{\ell} P^{(\ell)}_{{\bf p},i,m}$ one finds
\begin{equation}
\label{eq:coef-expn-Pnl}
\begin{aligned}
& \lambda_{i,n}^{(\ell)} = \frac{1}{A_{i,n}^{(\ell)}} \left\langle \left(\Delta_{\mu \nu} p^{\mu}_{i} p^{\nu}_{i} \right)^{\ell} \frac{E_{\mathbf{p},i}}{\tau_{R{\bf p},i}} G(E_{\mathbf{p},i}) P^{(\ell)}_{{\bf p},i,n} \right\rangle_{\widetilde{\mathrm{eq}}, i}.
\end{aligned}    
\end{equation}

Given Eq.~\eqref{eq:coef-expn-Pnl}, we compute the rank-2 tensor component of $\phi^{(1)}_{{\bf p},i}$. From the $\ell=2$ term in Eq.~\eqref{eq:expn-tensorial} and using the solution for the $\Phi$ components obtained in  Eq.~\eqref{eq:soln-Phis-l=2} one has,
\begin{equation}
\label{eq:F2-sum}
\begin{aligned}
& F^{(2)}_{\mathbf{p},i} =  \beta \sum_{n=0}^{\infty} \frac{1}{A_{i,n}^{(2)}}  \left\langle  \left(\Delta_{\mu \nu} p^{\mu}_{i} p^{\nu}_{i}\right)^{2} P^{(2)}_{{\bf p},i,n} \right\rangle_{\widetilde{\mathrm{eq}}, i}  P^{(2)}_{{\bf p},i,n}.
\end{aligned}
\end{equation}
The term inside the sum has the form of Eq.~\eqref{eq:coef-expn-Pnl} for $G(E_{\mathbf{p},i}) = \tau_{R{\bf p},i}/E_{\mathbf{p},i}$ and $\ell = 2$. Using expansion \eqref{eq:expn-Pnl}, one finds that
\begin{equation}
\label{eq:F2-summed}
\begin{aligned}
& F^{(2)}_{\mathbf{p},i} = \beta \frac{\tau_{R{\bf p},i}}{E_{\mathbf{p},i}}.
\end{aligned}
\end{equation}
%

An analogous procedure is employed for the scalar component, $F^{(0)}$. Nevertheless, it is important to point out that the procedure is more involved since the matching fixation procedure imposes an extra step in the computation, due to the presence of zero-modes in the collision term. Then, from Eqs.~\eqref{eq:expn-tensorial} and \eqref{eq:soln-Phis-l=0}, we have 
\begin{equation}
\label{eq:F0-sum}
\begin{aligned}
& F^{(0)}_{\mathbf{p},i} = \Phi_{i,0} P_{{\bf p},i,0}^{(0)} - \sum_{n = 1}^{\infty}  \frac{1}{A^{(0)}_{i,n}} \left\langle \mathcal{A}_{{\bf p},i} P^{(0)}_{{\bf p},i,n} \right\rangle_{\widetilde{\mathrm{eq}},i}  P_{{\bf p},i,n}^{(0)} \theta
\end{aligned}
\end{equation}
It is seen that the second term of the above equation can be re-expressed by considering $G(E_{\mathbf{p},i}) = (\tau_{R{\bf p},i}/E_{\mathbf{p},i})\mathcal{A}_{{\bf p},i}$ and $\ell = 0$ in Eqs.~\eqref{eq:expn-Pnl} and \eqref{eq:coef-expn-Pnl},
\begin{equation}
\begin{aligned}
&
\sum_{n = 1}^{\infty}  \frac{1}{A^{(0)}_{i,n}} \left\langle \mathcal{A}_{{\bf p},i} P^{(0)}_{{\bf p},i,n} \right\rangle_{\widetilde{\mathrm{eq}},i}  P_{{\bf p},i,n}^{(0)}
=
\frac{\tau_{R{\bf p},i}}{E_{\mathbf{p},i}}\mathcal{A}_{{\bf p},i} - \frac{1}{A^{(0)}_{i,0}} \left\langle \mathcal{A}_{{\bf p},i} P^{(0)}_{{\bf p},i,0} \right\rangle_{\widetilde{\mathrm{eq}},i}  P_{{\bf p},i,0}^{(0)},
\end{aligned}
\end{equation}
and, hence
\begin{equation}
\begin{aligned}
& F^{(0)}_{\mathbf{p},i} = 
\Phi_{i,0} P_{{\bf p},i,0}^{(0)}
+
\left(
-
\frac{\tau_{R{\bf p},i}}{E_{\mathbf{p},i}}\mathcal{A}_{{\bf p},i} 
+
\frac{1}{A^{(0)}_{i,0}} \left\langle \mathcal{A}_{{\bf p},i} P^{(0)}_{{\bf p},i,0} \right\rangle_{\widetilde{\mathrm{eq}},i}  P_{{\bf p},i,0}^{(0)} \right) \theta.
\end{aligned}    
\end{equation}
Thus, Eq.~\eqref{eq:phi-all-sol-scalar} is established. In Appendix \ref{sec:zero modes and m}, we obtain the solution for $\Phi_{i,0}$, which will completely define the above equation.

\section{Zero-modes and matching conditions} \label{sec:zero modes and m}

In this Appendix, we provide details for the inversion of the novel RTA collision matrix consistently with Landau matching conditions, from which we obtain the components $\Phi_{i,0}$ and $\Phi_{i,0}^{\mu}$, for $i = 1, \cdots, N_{\mathrm{spec}}$ of expansion \eqref{eq:expn-tensorial}. As discussed in the main text, from  Eq.~\eqref{eq:CRTAs} we obtain the constrains, 
\begin{subequations}
\label{eq:sys-0-mode-apn}
\begin{align}
&
\label{eq:sys-0-mode-l=0-apn}
\sum_{j} \mathcal{M}^{(0)}_{ij} \Phi_{j,0} = - \langle \mathcal{A}_{{\bf p},i} P_{{\bf p},i,0}^{(0)} \rangle \theta, \quad \sum_{j} \mathcal{M}^{(1)}_{ij}\Phi^{\mu}_{j,0} = 0,\\
\label{eq:ML-def-apn}
&
\mathcal{M}^{(\ell)}_{ij}
=
A^{(\ell)}_{i,0}\left( \delta_{ij}  - \frac{A^{(\ell)}_{j,0}}{ \sum_{k} A^{(\ell)}_{k,0} }\right), \quad \ell = 0,1,
\end{align}    
\end{subequations}
Naively, since the above expressions form linear systems of $N_{\rm spec}$ equations for the $N_{\rm spec}$ unknowns $\Phi_{i,0}, \ i = 1, \cdots, N_{\rm spec}$ (and $\Phi_{i,0}^{\mu}, \ i = 1, \cdots, N_{\rm spec}$, respectively), we would solve them by simply inverting the symmetric matrices $\mathcal{M}^{(0)}$ and $\mathcal{M}^{(1)}$. However, property \eqref{eq:lin-coll-csv-law}, that is also observed for the novel RTA collision term, implies that the matrices $\mathcal{M}^{(0)}_{ij}$, $\mathcal{M}^{(1)}_{ij}$ obey
\begin{equation}
\label{eq:prop-M0-M1}
\begin{aligned}
&
\sum_{i} \mathcal{M}^{(0)}_{ij} = 0 = \sum_{i} \mathcal{M}^{(1)}_{ij}, 
\end{aligned}    
\end{equation}
This implies that $\det \mathcal{M}^{(0)} = 0 = \det \mathcal{M}^{(1)}$, since the lines of these matrices are not linearly independent. It can also be concluded that $\mathcal{M}^{(\ell)}$, $\ell = 0,1$ has one vanishing eigenvalue whose associated eigenvector is $\propto (1,1,...,1)$. We also assume that no other eigenvalues are zero.

The above statements imply that the Jordan decomposition of 
$\mathcal{M}^{(\ell)}, \ \ell = 0,1$ is $\mathcal{M}^{(\ell)} = \mathds{U}^{(\ell)} \mathcal{D}^{(\ell)} (\mathds{U}^{(\ell)})^{-1}$, with $\mathcal{D}^{(\ell)} = \text{diag}(\lambda_{1}^{(\ell)} = 0, \lambda_{2}^{(\ell)}, ..., \lambda_{N_{\rm spec}}^{(\ell)})$, $\lambda_{i}^{(\ell)} \neq 0, \ \forall i = 2, \cdots, N_{\rm spec}$ and $\mathds{U}^{(\ell)}$ is the matrix basis transformation toward the eigenvalue basis, for $\ell = 0,1$. The matrix $\mathds{U}^{(\ell)}$ is unitary, $(\mathds{U}^{(\ell)})^{-1} = (\mathds{U}^{(\ell)})^{T}$, and the columns $\mathds{U}^{(\ell)}$ are formed by the normalized orthogonal (with respect to the canonical $\mathds{R}^{N_{\rm spec}}$ inner product) eigenvectors of $\mathcal{M}^{(\ell)}$. From the latter, we deduce that the first column of $\mathds{U}^{(\ell)}$ (and, by extension, the first line of  $(\mathds{U}^{(\ell)})^{-1}$) is $(N_{\rm spec}^{-1/2},N_{\rm spec}^{-1/2}, ..., N_{\rm spec}^{-1/2})^{T}$.

Equations \eqref{eq:sys-0-mode-l=0-apn} can be schematically written as $\mathcal{M}^{(\ell)} \vec{\Phi}^{(\ell)} = \vec{\mathcal{S}}^{(\ell)}$, $\ell = 0,1$, with $\vec{\mathcal{S}}^{(1)} = (0,\cdots, 0)^{T}$, $\vec{\mathcal{S}}^{(0)} = -\left(\langle \mathcal{A}_{{\bf p},1} P_{1,0}^{(0)} \rangle \theta, \cdots, \langle \mathcal{A}_{{\bf p},N_{\rm spec}} P_{N_{\rm spec},0}^{(0)} \rangle \theta \right)^{T}$. One may now multiply both sides from the left by $(\mathds{U}^{(\ell)})^{-1}$, which yields $  (\mathds{U}^{(\ell)})^{-1} \mathcal{M}^{(\ell)}\vec{\Phi}^{(\ell)} = \mathcal{D}^{(\ell)} (\mathds{U}^{(\ell)})^{-1} \vec{\Phi}^{(\ell)} = (\mathds{U}^{(\ell)})^{-1}\vec{\mathcal{S}}^{(\ell)}$. The first equation in this system is $\lambda^{(\ell)}_{1} \times \sum_{j} \Phi_{j} = 0 \times  \sum_{j} \Phi_{j} = \sum_{j} \mathcal{S}_{j}^{(\ell)}$. However, $\sum_{j} \mathcal{S}_{j}^{(\ell)} = 0$, which is trivial for $\ell = 1$ and follows from property \eqref{eq:AE-zero} for $\ell = 0$. Hence, due to the properties of the collision term, the linear system of equations \eqref{eq:sys-0-mode-apn} can not be solved for the $\Phi$-components.   

We reinstate the inversibility of the linear systems \eqref{eq:sys-0-mode-apn} by complementing the linear system defined with the novel RTA collision matrix with the constrain imposed by the matching conditions. This shall be performed by expressing one of the $\Phi$ variables (e.g.~$\Phi_{1,0}$ and $\Phi_{1,0}^{\mu}$) in terms of the remaining components ($\Phi_{i \neq 1,0}$ and $\Phi_{i \neq 1,0}^{\mu}$), substituting in the systems \eqref{eq:sys-0-mode-apn} and generating a new system of linear equations defined by a non-singular matrix. Indeed, from Eqs.~\eqref{eq:matching-kin} and \eqref{eq:phi-part-sol} we have, respectively,
\begin{subequations}
\label{eq:match-Phis}
\begin{align}
&
\label{eq:match-Phis-l=0}
\sum_{j} \beta J_{3,0}^{(j)} \Phi_{j,0} 
+
\sum_{j}
\left(-\left\langle \tau_{R,j}E_{{\bf p},j}  \mathcal{A}_{{\bf p},j} \right\rangle_{\widetilde{\mathrm{eq}},j}
+
\frac{1}{A^{(0)}_{j,0}} \left\langle  \mathcal{A}_{{\bf p},j} P^{(0)}_{{\bf p},j,0} \right\rangle_{\widetilde{\mathrm{eq}},j}  \beta J_{3,0}^{(j)} \right) \theta
\equiv 0,   
\\
&
\label{eq:match-Phis-l=1}
\sum_{j} J_{3,1}^{(j)} \Phi_{j,0}^{\mu} \equiv 0 .
\end{align}    
\end{subequations}
From Eq.~\eqref{eq:match-Phis-l=1}, we have $\Phi_{1,0}^{\mu} =  - \sum_{j = 2}^{N_{\rm spec}} (J_{3,1}^{(j)}/J_{3,1}^{(1)}) \Phi_{j,0}^{\mu}$, which when substituted in the second expressio in  Eq.~\eqref{eq:sys-0-mode-l=0-apn} yields $\sum_{j = 2}^{N_{\rm spec}} [\mathcal{M}^{(1)}_{ij} -  (J_{3,1}^{(j)}/J_{3,1}^{(1)}) \mathcal{M}^{(1)}_{i1}]\Phi^{\mu}_{j,0} = 0$, $i = 2, \cdots, N_{\rm spec}$. In general, the $(N_{\mathrm{spec}}-1) \times (N_{\mathrm{spec}}-1)$ matrix $[\mathcal{M}^{(1)}_{ij} -  (J_{3,1}^{(j)}/J_{3,1}^{(1)}) \mathcal{M}^{(1)}_{i1}], i,j = 2, \cdots, N_{\rm spec}$ is in general non-singular and, thus, since the above is a linear homogeneous system of equations for the $\Phi^{\mu}_{i,0}, i = 2, \cdots, N_{\mathrm{spec}}$, we estabilish that $\Phi^{\mu}_{i \neq 1,0}$ are zero and consequently that $\Phi^{\mu}_{i = 1,0}$. And thus, as expected for a system with vanishing chemical potential, the vector component of the deviation function is zero. 

An analogous procedure can be made using Eqs.~\eqref{eq:sys-0-mode-l=0-apn}, \eqref{eq:match-Phis-l=0} and the $\Phi_{j,0}$, components. Hence, taking Eq.~\eqref{eq:match-Phis-l=0} and solving for $\Phi_{1,0}$, we have 
\begin{equation}
\label{eq:Phi-1}
\begin{aligned}
&
\Phi_{1,0} = 
 - \sum_{j = 2}^{N_{\rm spec}} \frac{J_{3,0}^{(j)}}{J_{3,0}^{(1)}} \Phi_{j,0} + 
 \frac{\theta}{\beta J_{3,0}^{(1)}}
 \sum_{j} \left[  \left\langle  \tau_{R {\bf p},j} E_{{\bf p},j}  \mathcal{A}_{{\bf p},j}  \right\rangle_{\widetilde{\mathrm{eq}},j}
-
 \frac{1}{A_{j,0}^{(0)}} \beta J_{3,0}^{(j)} \left\langle \mathcal{A}_{{\bf p},j} P^{(0)}_{{\bf p},j,0} \right\rangle_{\widetilde{\mathrm{eq}},j}\right],
\end{aligned}    
\end{equation}
 that, when substituted in Eq.~\eqref{eq:sys-0-mode-l=0} leads to the new system of equations for the remaining components $\Phi_{i,0} , i = 2, \cdots, N_{\mathrm{spec}}$
\begin{subequations}
\label{eq:subsys-num-eff-2}
\begin{align}
&
\sum_{j = 2}^{N_{\rm spec}} \widetilde{M}_{ij} \Phi_{j,0} = - \langle \mathcal{A}_{{\bf p},i} P_{{\bf p},i,0}^{(0)} \rangle_{\widetilde{\mathrm{eq}},i} \theta 
-
 \frac{\mathcal{M}_{i1}^{(0)}}{\beta J_{3,0}^{(1)}} \theta
 \sum_{k} \left(  \left\langle  \tau_{R {\bf p},k} E_{{\bf p},k}  \mathcal{A}_{{\bf p},k}  \right\rangle_{\widetilde{\mathrm{eq}},k}
-
 \frac{1}{A_{k,0}^{(0)}} \beta J_{3,0}^{(k)} \left\langle \mathcal{A}_{{\bf p},k} P^{(0)}_{{\bf p},k,0} \right\rangle_{\widetilde{\mathrm{eq}},k}\right),
\\
&
\notag
 \quad i = 2, \cdots, N_{\mathrm{spec}}, \\
&
\widetilde{M}_{ij} \equiv  \mathcal{M}_{ij}^{(0)} - \frac{J_{3,0}^{(j)}}{J_{3,0}^{(1)}} \mathcal{M}_{i1}^{(0)}, \quad i,j = 2, \cdots, N_{\mathrm{spec}},
\end{align}    
\end{subequations}
In general, the $(N_{\mathrm{spec}}-1) \times (N_{\mathrm{spec}}-1)$ matrix 
is in general non-singular and, thus, allows for the determination of the $\Phi_{j,0}$, $j = 2, \cdots, N_{\mathrm{spec}}$ by simple inversion. Namely,
\begin{subequations}
\label{eq:phi-0-sols}
\begin{align}
\Phi_{i \neq 1,0} & = -\sum_{j \neq 1}(\widetilde{M}^{-1})_{ij} \left[ \langle \mathcal{A}_{{\bf p},j} P_{{\bf p},j,0}^{(0)} \rangle_{\widetilde{\mathrm{eq}},j} \theta 
+
 \frac{\mathcal{M}_{j1}^{(0)}}{\beta J_{3,0}^{(1)}} \theta
 \sum_{k} \left(  \left\langle  \tau_{R {\bf p},k} E_{{\bf p},k}  \mathcal{A}_{{\bf p},k}  \right\rangle_{\widetilde{\mathrm{eq}},k}
-
 \frac{1}{A_{k,0}^{(0)}} \left\langle  E_{{\bf p},k}^{2} P_{{\bf p},k,0}^{(0)} \right\rangle \left\langle \mathcal{A}_{{\bf p},k} P^{(0)}_{{\bf p},k,0} \right\rangle_{\widetilde{\mathrm{eq}},k}\right)\right]
\notag
\\
&
=
-\sum_{j \neq 1}(\widetilde{M}^{-1})_{ij} \langle \mathcal{A}_{{\bf p},j} P_{{\bf p},j,0}^{(0)} \rangle_{\widetilde{\mathrm{eq}},j} \theta 
+
\frac{\theta}{\beta \mathcal{J}_{3,0}}
 \sum_{k} \left(  \left\langle  \tau_{R {\bf p},k} E_{{\bf p},k}  \mathcal{A}_{{\bf p},k}  \right\rangle_{\widetilde{\mathrm{eq}},k}
-
 \frac{1}{A_{k,0}^{(0)}} \left\langle  E_{{\bf p},k}^{2} P_{{\bf p},k,0}^{(0)} \right\rangle_{\widetilde{\mathrm{eq}},k} \left\langle \mathcal{A}_{{\bf p},k} P^{(0)}_{{\bf p},k,0} \right\rangle_{\widetilde{\mathrm{eq}},k}\right)
 \notag
\\
&
\label{eq:phi-0-sols-a}
=
-
 \frac{1}{A_{i,0}^{(0)}}
 \langle \mathcal{A}_{{\bf p},i} P_{{\bf p},i,0}^{(0)} \rangle_{\widetilde{\mathrm{eq}},i}
+
\frac{\theta}{\beta \mathcal{J}_{3,0}}
 \sum_{k} \left\langle  \tau_{R {\bf p},k} E_{{\bf p},k}  \mathcal{A}_{{\bf p},k}  \right\rangle_{\widetilde{\mathrm{eq}},k},
\\
\Phi_{i = 1,0} & =- \theta \sum_{j \neq 1} \frac{J_{3,0}^{(j)}}{J_{3,0}^{(1)}}
\left[-
 \frac{1}{A_{j,0}^{(0)}}
 \langle \mathcal{A}_{{\bf p},j} P_{{\bf p},j,0}^{(0)} \rangle_{\widetilde{\mathrm{eq}},j}
+
\frac{\theta}{\beta \mathcal{J}_{3,0}}
 \sum_{k} \left\langle  \tau_{R {\bf p},k} E_{{\bf p},k}  \mathcal{A}_{{\bf p},k}  \right\rangle_{\widetilde{\mathrm{eq}},k}\right] 
\notag
\\
&
+ 
 \frac{\theta}{\beta J_{3,0}^{(1)}}
 \sum_{j} \left(  \left\langle  \tau_{R {\bf p},j} E_{{\bf p},j}  \mathcal{A}_{{\bf p},j}  \right\rangle_{\widetilde{\mathrm{eq}},j}
-
 \frac{1}{A_{j,0}^{(0)}} \beta J_{3,0}^{(j)} \left\langle \mathcal{A}_{{\bf p},j} P^{(0)}_{{\bf p},j,0} \right\rangle_{\widetilde{\mathrm{eq}},j}\right),
\notag
\\
&
=
-
 \frac{1}{A_{1,0}^{(0)}}
 \langle \mathcal{A}_{{\bf p},1} P_{{\bf p},1,0}^{(0)} \rangle_{\widetilde{\mathrm{eq}},1}
+
\frac{\theta}{\beta \mathcal{J}_{3,0}}
 \sum_{k} \left\langle  \tau_{R {\bf p},k} E_{{\bf p},k}  \mathcal{A}_{{\bf p},k}  \right\rangle_{\widetilde{\mathrm{eq}},k},
\label{eq:phi-0-sols-b}
\end{align}    
\end{subequations}
where, in the second equality of Eq.~\eqref{eq:phi-0-sols-a}, we have evoked the following property of the $\widetilde{M}^{-1}$ matrix,
\begin{equation}
\label{eq:M-M-1-prop}
\begin{aligned}
\sum_{j \neq 1}(\widetilde{M}^{-1})_{ij} \mathcal{M}^{(0)}_{j1} 
=
- \frac{J_{3,0}^{(1)}}{\mathcal{J}_{3,0}}, \quad
i = 2, \cdots, N_{\mathrm{spec}},
\end{aligned}    
\end{equation}
which can be derived directly from the definition of $\widetilde{M}$ in Eq.~\eqref{eq:subsys-num-eff-2} (see  section \ref{apn:prop-MM1} below). In the third equality of Eq.~\eqref{eq:phi-0-sols-a}, we used the fact that $P^{(0)}_{{\bf p}, i,0} = \beta E_{{\bf p},i}$, and that the inverse $\widetilde{M}^{-1}$ can be determined analytically to be (see subsection \ref{apn:prop-MM1})
\begin{equation}
\label{eq:inv-M-analyt}
\begin{aligned}
&
(\widetilde{M}^{-1})_{ij}
=
\frac{1}{A^{(0)}_{i,0}} \delta_{ij}
+
\frac{1}{A^{(0)}_{j,0}} \frac{1}{A^{(0)}_{1,0}} 
\frac{J_{3,0}^{(1)}}{\mathcal{J}_{3,0}} \left( A^{(0)}_{j,0}
-
\frac{J_{3,0}^{(j)}}{J_{3,0}^{(1)}} 
 A^{(0)}_{1,0}
 \right), 
\end{aligned}    
\end{equation}
which significantly simplifies the expression. The solution for $\Phi_{i = 1,0}$, Eq.~\eqref{eq:phi-0-sols-b}, is obtained by inserting Eq.~\eqref{eq:phi-0-sols-a} in Eq.~\eqref{eq:Phi-1}. From the first to the second equality in Eq.~\eqref{eq:phi-0-sols-b}, the first term in square brackets partially cancels the second term in round brackets and second term in square brackets partially cancels the first term in round brackets. This allows for the complete definition of the scalar sector of the deviation function, $\left. \Hat{\phi}^{(1)}_{{\bf p},i} \right\vert_{\ell = 0}$. We can readily notice that even though the separate expressions for $\Phi_{i = 1,0}$ and $\Phi_{i \neq 1,0}$ are at first necessary in the derivation, the final expression can be put in the unified form 
\begin{equation}
\label{eq:phi-0-sols-unif-apn}
\begin{aligned}
&
\Phi_{i,0}
=
-
 \frac{\theta}{A_{i,0}^{(0)}}
 \langle \mathcal{A}_{{\bf p},i} P_{{\bf p},i,0}^{(0)} \rangle_{\widetilde{\mathrm{eq}},i}
+
\frac{\theta}{\beta \mathcal{J}_{3,0}}
 \sum_{k} \left\langle  \tau_{R {\bf p},k} E_{{\bf p},k}  \mathcal{A}_{{\bf p},k}  \right\rangle_{\widetilde{\mathrm{eq}},k},
\quad i = 1, \cdots, N_{\rm spec}. 
\end{aligned}    
\end{equation}
This is consistent, since the choice of `particle species 1' is essentially arbitrary, and hence the results could not depend on which specific species is labeled as `1'. We note that the first term of the equation for $\Phi_{i,0}$ above exactly cancels the second term in brackets in Eq.~\eqref{eq:phi-all-sol-scalar} for 
$\left. \Hat{\phi}^{(1)}_{{\bf p},i} \right\vert_{\ell = 0}$.

\subsection{Properties of the matrix $\widetilde{M}$}
\label{apn:prop-MM1}

In the above section, we derived expressions for the $\Phi$ scalar expansion components related to the zero modes of the collision term. From the collisional matrix $\mathcal{M}^{(0)}$ and the matching conditions \eqref{eq:matching-kin}, a non-singular matrix $\widetilde{M}$ is defined (see Eq.~\eqref{eq:subsys-num-eff-2}). The inversion of the linear system defined with $\widetilde{M}$ leads to Eqs.~\eqref{eq:phi-0-sols}. The matrix $\widetilde{M}$ can be expressed as 
\begin{equation}
\begin{aligned}
&
\widetilde{M}_{ij} \equiv  \mathcal{M}_{ij}^{(0)} - \frac{J_{3,0}^{(j)}}{J_{3,0}^{(1)}} \mathcal{M}_{i1}^{(0)}
=
A^{(0)}_{i,0} \delta_{ij}  - \frac{A^{(0)}_{i,0}}{ \sum_{k} A^{(0)}_{k,0} } \left( A^{(0)}_{j,0}
-
\frac{J_{3,0}^{(j)}}{J_{3,0}^{(1)}} 
 A^{(0)}_{1,0}
 \right)
 \equiv
a_{i}\delta_{ij} - a_{i} b_{j} 
, \quad i,j = 2, \cdots, N_{\mathrm{spec}}
\end{aligned}    
\end{equation}
where $a_{i} = A^{(0)}_{i,0}$ and $b_{j} = (1/ \sum_{k} A^{(0)}_{k,0}) \left( A^{(0)}_{j,0}
-
\frac{J_{3,0}^{(j)}}{J_{3,0}^{(1)}} 
 A^{(0)}_{1,0}
 \right)$.
By inspection, one can derive that the specific form of the matrix allows for analytical expression of the inverse
\begin{equation}
\begin{aligned}
&
(\widetilde{M}^{-1})_{ij}
=
\frac{1}{a_{i}} \delta_{ij}
+
\frac{b_{j}}{a_{j}} \frac{1}{1-\sum_{n} b_{n}}.
\end{aligned}    
\end{equation}
As a matter of fact one can check that indeed
\begin{subequations}
\begin{align}
\sum_{k}(\widetilde{M}^{-1})_{ik} \widetilde{M}_{kj}
&=
\sum_{k} \left(\frac{1}{a_{i}} \delta_{ik}
+
\frac{b_{k}}{a_{k}} \frac{1}{1-\sum_{n} b_{n}} \right) (a_{k}\delta_{kj} - a_{k} b_{j})
\notag
\\
&
=
\sum_{k} \left(\frac{a_{k}}{a_{i}}\delta_{ik} \delta_{kj} 
+
\frac{b_{k}}{1-\sum_{n} b_{n}} \delta_{kj}
-  b_{j}
\frac{a_{k}}{a_{i}} \delta_{ik}
-  \frac{b_{j} b_{k}}{1-\sum_{n} b_{n}}
\right) 
=
\delta_{ij},
\\
\sum_{k} \widetilde{M}_{ik} (\widetilde{M}^{-1})_{kj}
&= 
\sum_{k} (a_{i}\delta_{ik} - a_{i} b_{k}) \left(\frac{1}{a_{k}} \delta_{kj}
+
\frac{b_{j}}{a_{j}} \frac{1}{1-\sum_{n} b_{n}} \right) 
\notag
\\
&
=
\sum_{k}  \left(
\frac{a_{i}}{a_{k}} \delta_{ik} \delta_{kj} - b_{k} \frac{a_{i}}{a_{k}} \delta_{kj}
+
a_{i}\delta_{ik}\frac{b_{j}}{a_{j}} \frac{1}{1-\sum_{n} b_{n}}
-
a_{i} b_{k}
\frac{b_{j}}{a_{j}} \frac{1}{1-\sum_{n} b_{n}}
\right) 
= 
\delta_{ij}.
\end{align}    
\end{subequations}
And hence, we have that
\begin{equation}
\label{eq:M-1-analyt}
\begin{aligned}
&
(\widetilde{M}^{-1})_{ij}
=
\frac{1}{A^{(0)}_{i,0}} \delta_{ij}
+
\frac{1}{A^{(0)}_{j,0}}  \left( A^{(0)}_{j,0}
-
\frac{J_{3,0}^{(j)}}{J_{3,0}^{(1)}} 
 A^{(0)}_{1,0}
 \right) \frac{1}{A^{(0)}_{1,0}} 
\frac{J_{3,0}^{(1)}}{\mathcal{J}_{3,0}}.
\end{aligned}    
\end{equation}

In the final stage of the derivation in main part of the present Appendix, property \eqref{eq:M-M-1-prop} is employed. This property follows from the definitions of $\widetilde{M}$ and $\mathcal{M}^{(0)}$ and we derive it below. Namely, by the definition of an inverse matrix, we have
\begin{equation}
\begin{aligned}
&
\sum_{j \neq 1}(\widetilde{M}^{-1})_{ij}\widetilde{M}_{jk} = \delta_{ik}, \quad i,k = 2, \cdots, N_{\mathrm{spec}},  
\\
&
\Rightarrow
\sum_{j \neq 1}(\widetilde{M}^{-1})_{ij}\left(\mathcal{M}_{jk}^{(0)} - \frac{J_{3,0}^{(k)}}{J_{3,0}^{(1)}} \mathcal{M}_{j1}^{(0)} \right) = \delta_{ik}.
\end{aligned}    
\end{equation}
Performing a summation $\sum_{k \neq 1}$, we have 
\begin{equation}
\begin{aligned}
&
-\left(1 + \frac{\sum_{k \neq 1} J_{3,0}^{(k)}}{J_{3,0}^{(1)}}\right)\sum_{j \neq 1}(\widetilde{M}^{-1})_{ij}\mathcal{M}_{j1}^{(0)} = 1,
\end{aligned}    
\end{equation}
where we employed property \eqref{eq:prop-M0-M1} in the form $\sum_{k \neq 1} \mathcal{M}^{(0)}_{jk} = - \mathcal{M}^{(0)}_{j1}$. Simple algebraic manipulations lead to property \eqref{eq:M-M-1-prop}. This property can also be checked with the analytical expression \eqref{eq:M-1-analyt}.

%


\bibliographystyle{apsrev4-1}
\bibliography{liografia}

\end{document}